\RequirePackage{fix-cm}
\documentclass[smallextended]{svjour3}       
\smartqed  
\usepackage{mathptmx} 
\usepackage{graphicx} 
\usepackage{amstext} 
\usepackage[top=25mm,bottom=37mm,left=20mm,right=20mm,columnsep=10mm]{geometry} 
\usepackage{color} 
\definecolor{myblu}{rgb}{0.1,0.1,0.5}
\usepackage{hyperref} 
\hypersetup{colorlinks=true,urlcolor=blue,linkcolor=black,citecolor=black} 
\usepackage{secdot} 
\usepackage{amsmath}
\usepackage{subfig}
\usepackage{psfrag}
\usepackage{float}
\usepackage[numbers]{natbib}
\usepackage{pifont}
\usepackage{relsize}
\usepackage{graphicx}
\usepackage{tabularx}
\usepackage{array}
\usepackage{multirow}
\usepackage{booktabs}
\usepackage{systeme}
\usepackage{scalerel,amssymb}
\begin{document}

\title{Passive flow-separation control in a dual-bell rocket nozzle}


\author{M. Cimini         \and
        E. Martelli       \and
	M. Bernardini
}


\institute{M. Cimini \at
              Sapienza University of Rome\\ 
	      Via Eudossiana 18, Rome, 00184, Italy\\
              \email{matteo.cimini@uniroma1.it}           
           \and
           E. Martelli \at
              University of Campania "L. Vanvitelli"\\
	      Caserta, 81100, Italy\\
	      \email{emanuele.martelli@unicampania.it}
	   \and
	   M. Bernardini (corresponding author) \at
	      Sapienza University of Rome\\ 
              Via Eudossiana 18, Rome, 00184, Italy\\
              \email{matteo.bernardini@uniroma1.it}
}

\date{Received: date / Accepted: date}

\maketitle

\begin{abstract}
A delayed detached eddy simulation of a sub-scale cold-gas dual-bell nozzle flow at high
Reynolds number and in sea-level mode is carried out at nozzle pressure
ratio NPR=45.7. In this regime the over-expanded flow exhibits a symmetric
and controlled flow separation at the inflection point, that is the junction 
between the two bells, leading to the
generation of a low content of aerodynamic side loads with respect to conventional
bell nozzles. The nozzle wall-pressure signature is analysed in
the frequency domain and compared with the experimental data available
in the literature for the same geometry and flow conditions. The Fourier
spectra in time and space (azimuthal wavenumber) show the presence of a persistent tone
associated to the symmetric shock movement. Asymmetric modes are only slightly
excited by the shock and the turbulent structures. The low mean value of
the side-loads magnitude is in good agreement with the experiments and
confirms that the inflection point dampens the aero-acoustic interaction between the separation-shock 
and the detached shear layer.

\end{abstract}


\section{Introduction}
\label{intro}

\begin{figure}
 \centering
  \includegraphics[height=9.3cm,angle=270,clip]{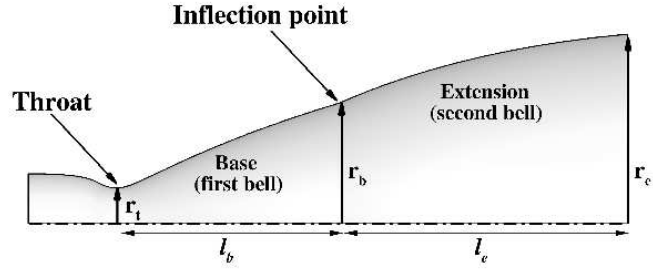} 
  \caption{Schematic of the dual-bell nozzle geometry.}
 \label{fig:db_model}
\end{figure}

The dual-bell nozzle is a kind of one-step altitude compensating
nozzle that may represent a possible alternative to replace conventional bell
nozzles in future launcher first-stage engines. 
The main feature of this advanced concept is the
particular shape of the divergent section, composed by two bells, namely the base
and the extension, separated by an inflection point, as shown in Fig.~\ref{fig:db_model}.
It has two main operating conditions: i) sea-level mode and ii) high-altitude mode.
In the first mode, it operates with a low area ratio, with a controlled and
symmetrical flow separation at the inflection point, thus avoiding the onset of 
dangerous side loads than can induce structural damages to the engine. 
According to Schmucker~\citep{schmucker}, side loads are caused by a non-symmetrical 
shock-induced flow separation and the degree of this non-symmetric behaviour is proportional 
to the inverse of the wall-pressure gradient magnitude. The inflection point in the dual-bell contour 
induces an infinite wall-pressure gradient (in the inviscid sense) thus zeroing the side-load magnitude.
In the high-altitude
mode, the flow attaches to the extension and the nozzle works with a higher area ratio,
thus increasing the thrust coefficient.
The parameter governing the transition between the two operating modes
is the nozzle pressure ratio (NPR), i.e. the ratio between the
chamber and the ambient pressure $p_0/p_a$.

The dual-bell nozzle was first proposed by Cowles and Foster~\citep{cowles1949} in 1949 and 
all the studies carried out since then highlighted three main critical issues: 
i) the transition between the two operating modes, 
ii) the detached flow unsteadiness in sea-level mode and iii) hot flow
behaviour and cooling effects.
The first experimental test campaign was performed by Horn and Fisher~\citep{horn1993} 
in 1993 using cold-gas sub-scale dual-bell nozzles. They found that both the constant-pressure 
and increasing-pressure extension exhibited good transition characteristics 
with a transition time of the order of 30 ms.
Several numerical studies~\citep{wong2002,nasuti2005} confirmed a quick transition 
behaviour. In particular, \citet{nasuti2005,nasuti2002numerical} found that the dual-bell 
extension with a positive pressure gradient can ensure
a transition faster than the one with constant pressure profile (CP). On the other hand
this profile can ensure higher thrust performances and for this reason a trade off is necessary.
%
%
%
After the experiments by Horn and Fisher, the transition phenomenon 
was studied through an extensive test campaign 
conducted at the German Aerospace Research Center (DLR)~\citep{genin2009,genin2010}, 
that investigated the effect of 
different geometrical parameters on the transition process and on the side-loads 
generation in cold-gas sub-scale nozzles.
They found that both modes (sea-level and high-altitude) were associated to a level of side loads 
lower than the ones suffered by a comparable truncated ideal contour (TIC) nozzle. Nevertheless, the 
transition was characterized by a short-time high-peak side load, that could jeopardize the nozzle structure.

A peculiar characteristics of the switching between the two operating modes was indicated by~\citet{nasuti2005}, 
who found that at the beginning of the transition the separation front moves into an  
inflection region, located immediately downstream of the inflection point
and characterized by a negative wall-pressure gradient (due to viscosity) like conventional nozzles.
As a consequence the dual bell can start to experience non-axial forces before the full transition takes place. 
The length of the inflection region was found to be of the order of the throat radius and, as shown 
by~\citet{martelli2007}, it depends on the boundary-layer thickness at the end of the base, 
the Prandtl-Meyer expansion fan at the inflection point and the wall-pressure gradient of the 
extension. \citet{geninstark2011} confirmed experimentally the existence of the inflection region and
that during the transition of the separation point in this zone the level of the side loads is similar
to that suffered by a conventional nozzle. 
Verma et al.~\citep{verma2015} studied  the flow unsteadiness when 
the separation front was located in the inflection region by analyzing the spectral content of the wall-pressure 
signature. They related the onset of side loads during the transition in the inflection region to the high 
level of unsteadiness suffered by the flow in this regime.
Another important aspect considered in literature was the effect of the launcher-base flow 
on the nozzle internal flow behaviour. 
In this context the investigation of the complex interaction of a dual-bell exhaust flow with 
the unsteady external flow is of particular interest, 
especially when the nozzle works in sea-level mode. 
As observed and demonstrated~\citep{perigo2003,torngren2002,wong2005} an external pressure 
fluctuation can cause an internal amplification of the pressure oscillations, which can be 
dangerous for the side-loads generation. \citet{verma2014} suggested that the external unsteady
perturbation could lead to a transition/re-transition cycle, generally known as flip-flop effect. 
~\citet{loosen2019ftc} numerically studied the interference of a turbulent wake coming from a
generic space launcher with a dual-bell nozzle exhaust flow in sea-level mode. It was found
that in supersonic freestream conditions, the presence of the outer flow leads to a premature transition,
thus reducing the transition NPR. 

Most of the studies available in literature refer to cold-gas flow investigations. In real flight 
conditions however the dual-bell nozzle would work with hot exhaust flow and a film
cooling technique could be used to protect the nozzle structure to thermal fatigue. \citet{genin2013hot}
performed one of the first experimental test campaign in order to study the nozzle behaviour
with inert hot air flow. The presence of the inflection point induces thermal loads during the
transition process and the generated heat flux raises in the inflection region. 
\citet{martelli2009hot} studied the effect of a film cooling injected in the base 
near the inflection point by means of numerical simulations. The emerging results highlight the 
efficiency of the film which is strongly influenced by the expansion fan at the inflection point. 
Unfortunately, from the study also emerges that the secondary gas increases the inflection region 
extension, leading to an increased risk of side-loads generation.

From the analysis of the available literature, it is clear that the development
of side loads inside the nozzle is one of the main critical
aspects that must be taken under consideration for the development and the realisation of a
dual-bell nozzle. 
This manuscript focuses on the numerical investigation of the development of 
lateral loads inside an over-expanded dual-bell nozzle working in the sea-level mode. 
From a numerical point of view, the best approach to resolve the flow-separation dynamics
and the shock-wave/boundary-layer interaction would be the large eddy simulation (LES) technique.
But performing a wall-resolved LES of this
high Reynolds numbers flows ($Re \approx 10^7$), requires an impractically computational
effort. On the other hand, modeling
the global effect of the turbulent scales as done in the URANS approach could suppress
the important flow processes leading to the formation of the aerodynamic unsteady loads.
In this scenario, a possible and valid choice is represented by the use of 
a hybrid RANS/LES methodology~\citep{Spalart2006,Spalart2009,frohlich2008,sandberg2014}, 
whose rationale is to simulate
attached boundary layers in RANS mode, while separated
shear layers and turbulent recirculating zones are solved by the LES mode.
Among the different methods, in this work we chose to adopt 
the detached eddy simulation (DES) technique~\citep{spalart97} and, 
in this framework, few test cases of separated dual-bell nozzle flows are reported in the literature.
In particular, \citet{proschanka2012} performed a numerical study of 
a cold-gas dual-bell nozzle in sea-level mode, founding three types
of pressure fluctuations: one symmetric and two asymmetric, the latter being associated
with side-loads generation.

In this work, a delayed detached eddy simulation (DDES) of a sub-scale cold-gas
dual-bell nozzle in the first operating mode is carried out.
The geometry is inspired by the experimental work of \citet{verma2015}
and it is characterized by a high nozzle-exit Mach number, closer to 
a real application than the
dual bell used by \citet{proschanka2012}.
The selected simulated NPR is experimentally characterized by a very low level of side loads, 
since the separation front is located at the inflection point, at the very beginning 
of the inflection region (that is with a very high value in modulus of the wall-pressure gradient).
The present analysis is focused on the investigation of the wall-pressure signature and
its spectral content, in particular in the azimuthal wave-number frequency plane, in order
to find a correlation between the forced flow-separation and the energy level of the non-symmetrical
azimuthal mode. Indeed, it has been argued by~\citet{Jaunet2017} and~\citet{martelli2019flow} that 
the side loads could be originated by an aeroacoustic feed-back loop (screech-like phenomenon) 
involving the shock-cell structure and the detached shear layer. The inflection point 
should change the receptivity of the separation-shock, thus altering the feed-back loop. 
To this purpose, the side-loads content inside the dual bell is analysed and compared
with experimental results and with the  TIC nozzle studied by ~\citet{martelli2019flow} with
the same numerical methodology.
The paper is organised as follows. First, the methodological approach and the
computational setup are presented in Sec.~\ref{sec:comp} and~\ref{sec:setup}. After providing an
overview of the flowfield organisation in Sec.~\ref{sec:flow_org}, the main features
of the wall-pressure signature are analysed in Sec.~\ref{sec:wallp} and Sec.\ref{sec:azimuthal} 
by means of spectral analysis (through Fourier transform in time and wavenumber-frequency
decomposition) and compared with experimental data. Then the aerodynamic side 
loads generated by the nozzle are inspected and compared with experiments and
numerical data of a TIC nozzle in Sec.~\ref{sec:sl}. Conclusions are finally given in 
Sec.~\ref{sec:conclusions}.

\section{Methodology}
\label{sec:comp}

In the present work we consider the governing equations for a compressible, viscous and heat-conducting
gas, that in three dimensions can be written in conservation form as

\begin{equation}
 \left .
 \begin{aligned}
   \frac{\partial \rho}{\partial t} + \frac{\partial (\rho \, u_j)}{\partial x_j}
     & = 0, \\
   \frac{\partial (\rho \, u_i)}{\partial t} + \frac{\partial (\rho \, u_i u_j)}{\partial x_j} +
       \frac{\partial p}{\partial x_i} - \frac{\partial \tau_{ij}}{\partial x_j} & = 0, \\
   \frac{\partial (\rho \, E)}{\partial t} + \frac{\partial (\rho \, E u_j + p u_j)}{\partial x_j}
     - \frac{\partial (\tau_{ij} u_i - q_j)}{\partial x_j} & = 0, \label{eq:ns}
 \end{aligned}
 \right .
\end{equation}

where $\rho$ is the density of the flow, $u_i$ denotes the velocity component in the $i$-th
coordinate direction ($i=1,2,3$), $E$ is the total energy per unit mass and
$p$ is the thermodynamic pressure.
$\tau_{ij}$ and $q_j$ denote the total stress tensor and total heat flux, that are given by
the sum of a viscous and a turbulent contribution, according to
\begin{equation}
 \label{eq:stress}
 \tau_{ij} = 2 \, \rho \left ( \nu + \nu_t \right ) S^*_{ij} \qquad
 q_j = -\rho \, c_p \left ( \frac{\nu}{\mathrm{Pr}} + \frac{\nu_t}{\mathrm{Pr}_t} \right ) \frac{\partial T}{\partial x_j}
\end{equation}
where the Boussinesq hypothesis is applied through the introduction of the
eddy viscosity $\nu_t$ (defined by the turbulence model), $S^*_{ij}$ is the traceless
strain-rate tensor and $\nu$ the kinematic viscosity,
depending on temperature $T$ through the Sutherland's law. The molecular and turbulent Prandtl numbers
$\mathrm{Pr}$ and $\mathrm{Pr}_t$ are considered constant and equal to 0.72 and 0.9, respectively.

The numerical methodology employed in this work is the delayed detached eddy simulation~\citep{spalart06},
an efficient and powerful hybrid RANS/LES technique well suited to capture the unsteadiness
of high-Reynolds number turbulent flows. The current implementation is based
on the Spalart-Allmaras (SA) turbulence model,
which solves a transport equation for a pseudo eddy viscosity $\tilde{\nu}$

\begin{equation}
 \label{eq:sa}
         \frac{\partial (\rho \tilde{\nu})}{\partial t} + \frac{\partial (\rho \, \tilde{\nu} \,u_j)}{\partial x_j} = c_{b1} \tilde{S} \rho \tilde{\nu} +
        \frac{1}{\sigma} \left [ \frac{\partial}{\partial x_j} \left [ \left ( \rho \nu + \rho \tilde {\nu} \right ) \frac{\partial \tilde{\nu}}{\partial x_j} \right ] +
         +c_{b2} \, \rho \left ( \frac{\partial \tilde{\nu}}{\partial x_j} \right )^2 \right ] 
         -c_{w1}  f_w \rho \left ( \frac{\tilde{\nu}}{\tilde{d}} \right )^2 ,
\end{equation}

where $\tilde{d}$ is the model length scale, $f_w$ is a near-wall damping
function, $\tilde{S}$ is a modified vorticity magnitude, and
$\sigma, c_{b1}, c_{b2}, c_{w1}$ are model constants.
The pseudo eddy viscosity $\tilde{\nu}$ is directly linked to the eddy viscosity $\nu_t$
through $\nu_t = \tilde{\nu} \, f_{v1}$, where the correction function $f_{v1}$ is used to guarantee the correct
near-wall boundary-layer behaviour.
In the DDES approach the turbulence model automatically switches between a pure RANS mode,
active in flow regions with attached boundary layers, to a pure LES mode, active in flow regions
detached from the wall, where the computation can directly resolve the large scale, energy
containing eddies.
This objective is achieved by defining the length-scale $\tilde{d}$ in~\eqref{eq:sa} as
\begin{equation}
 \tilde{d} = d_w - f_d \, \textrm{max} \left (0, d_w-C_{DES} \, \Delta \right),
\end{equation}
where $d_w$ is the distance from the nearest wall, $\Delta$ is the subgrid
length-scale that controls the wavelengths resolved in LES mode and $C_{DES}$
is a calibration constant set equal to 0.65 in the original model.  On the basis of previous
calibration studies performed on DDES of a sub-scale rocket nozzle~\cite{martelli2019,martelli2019flow}
the constant $C_{DES}$ has been set equal to 0.20 and the  function $f_d$ here employed is

\begin{equation}
 f_d = 1-\tanh{\left [ \left ( 16 r_d \right )^3 \right ]}, \qquad r_d = \frac{\tilde{\nu}}{k^2 \, d_w^2 \, \sqrt{U_{i,j} U_{i,j}}},
\end{equation}

where $U_{i,j}$ is the velocity gradient and $k$ the von Karman constant.
The $f_d$ function, that is built in such a way that its value is $0$ in boundary layers
and $1$ in LES regions, represents the main difference between the DDES strategy and the original DES
approach~\citep{spalart97}, denoted as DES97. It guarantees that attached boundary layers are always treated in RANS mode,
even in the case of extremely fine grids, thus allowing to alleviate the
well-known phenomenon of modeled stress depletion, which in turn can
lead to grid-induced separation~\cite{spalart06}.
The sub-grid length scale is specified according to the formulation proposed by~\citet{Deck2012},
and it depends on the flow itself, through $f_d$ as

\begin{equation}
 \label{eq:delta}
 \Delta = \frac{1}{2}
 \left [
 \left (1+\frac{f_d-f_{d0}}{|f_d-f_{d0}|} \right ) \, \Delta_{\textrm{max}} +
 \left (1-\frac{f_d-f_{d0}}{|f_d-f_{d0}|} \right ) \, \Delta_{\textrm{vol}}
 \right ],
\end{equation}

with $f_{d0} = 0.8$, $\Delta_{\mathrm{max}} = \max (\Delta x, \Delta y, \Delta z)$
and $\Delta_{\mathrm{vol}} = (\Delta x \cdot \Delta y \cdot \Delta z)^{1/3}$.
The main idea of this formulation is to take advantage of the $f_d$ function to switch
between $\Delta_{max}$, needed to shield the boundary layer, and $\Delta_{\mathrm{vol}}$,
needed to ensure a rapid destruction of modelled viscosity to unlock the
Kelvin-Helmholtz instability and accelerate the passage to resolved turbulence in
the separated shear layer.

\section{Computational setup}
\label{sec:setup}

\subsection{Flow solver}

The simulations have been performed by means of an in-house, compressible,
finite-volume, structured solver, widely employed in the past
to investigate the dynamics of turbulent, separated flows in transonic and
supersonic rocket nozzles, involving complex shock-waves/boundary-layer interactions~\citep{martelli2017,
martelli2019,memmolo2018}.
In the flow regions away from the shock, the spatial discretization consists
of a centered, second-order, energy consistent scheme,
that makes the numerical method extremely robust without the addition of numerical dissipation
~\citep{Pirozzoli2011}. This feature is particularly useful in the flow regions
treated in LES mode, where in addition to the molecular, the only relevant viscosity
should be that provided by the turbulence model.
Strong compressions in the flow are identified by means of
the Ducros shock sensor~\citep{ducrosetal99}, that is used to
switch the discretization of the convective terms of the governing equations
to third-order Weighted Essentially Non Oscillatory reconstructions
for cell-faces flow variables. The viscous fluxes are evaluated through compact,
second-order central-difference approximations.
A low-storage, third-order Runge-Kutta algorithm~\citep{Bernardini2009}
is used for time advancement of the semi-discretized ODEs' system.
The code is written in Fortran 90, it uses domain decomposition and it fully exploits
the message passing interface (MPI) paradigm for the parallelism.

\subsection{Test case description}

\begin{figure}
  \centering
(a)\includegraphics[height=7.0cm,angle=270,clip]{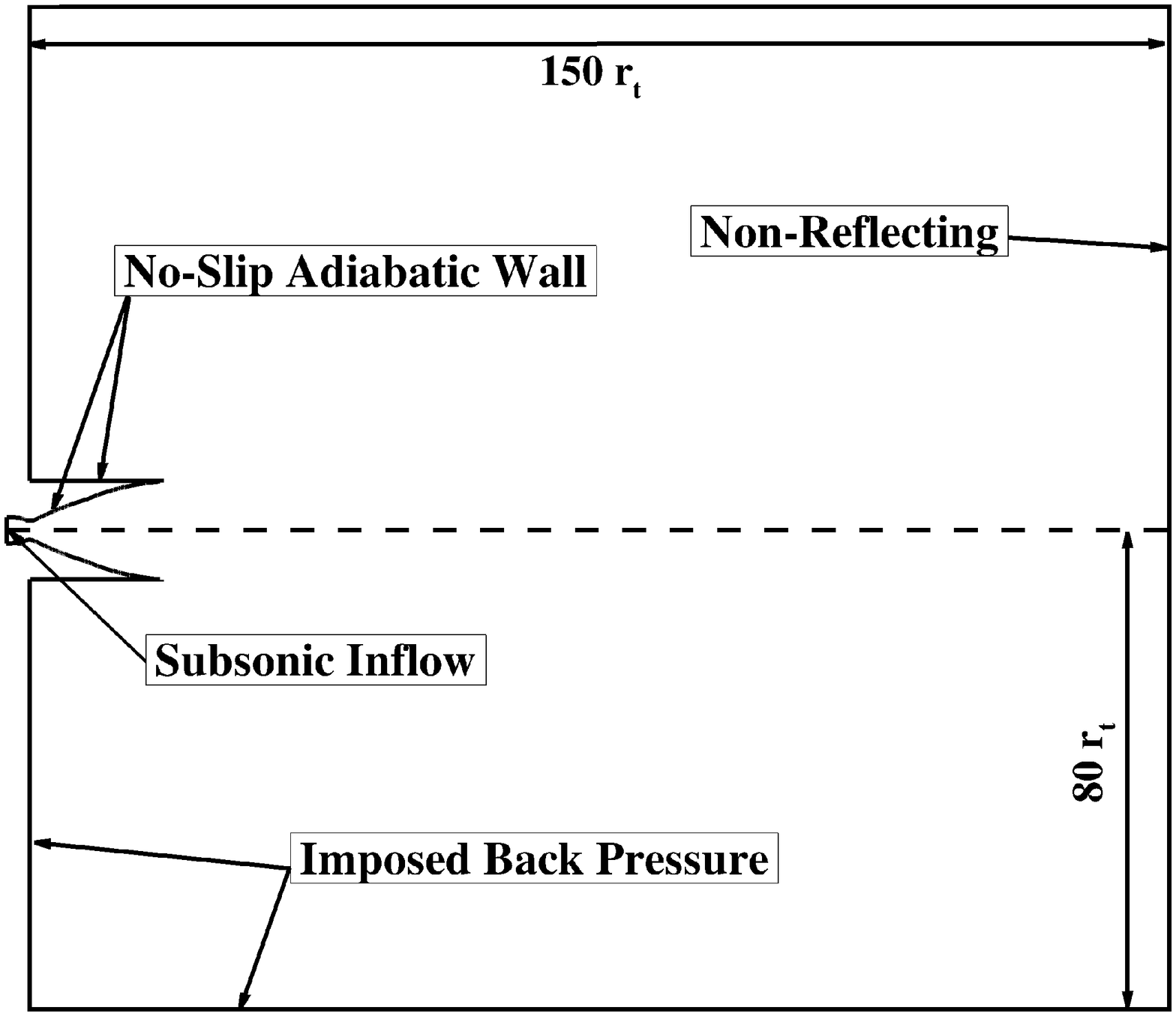} \hskip 1em
(b)\includegraphics[height=7.0cm,angle=270,clip]{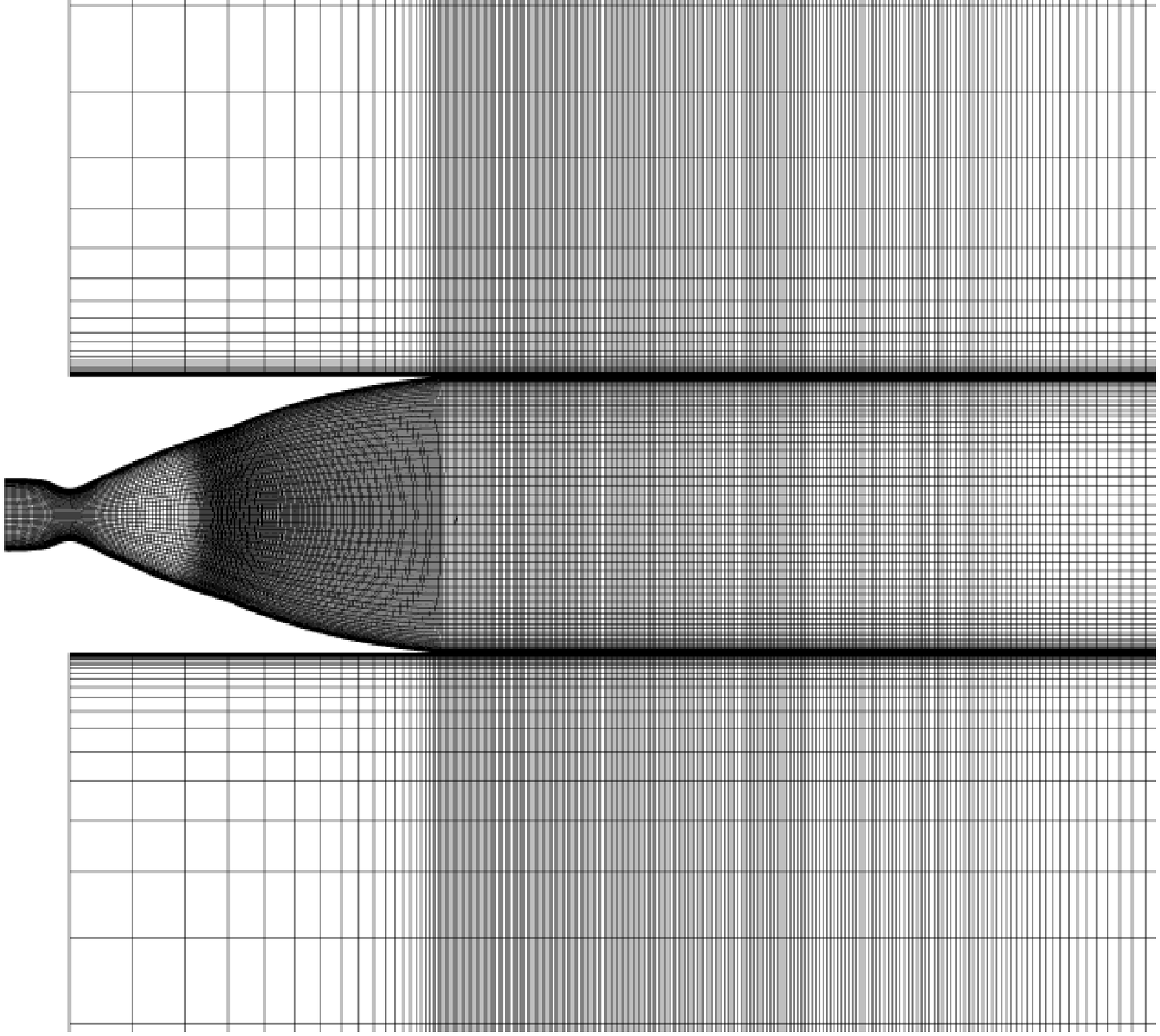} 
   \caption{Schematic of (a) the computational domain and (b) the computational grid adopted 
            for DDES in the x-y plane (only one every 8th grid nodes is shown).}
  \label{fig:mesh}
 \end{figure}


The present investigation is carried out on the dual-bell geometry experimentally
tested at DLR~\citep{verma2015},
characterized by a throat radius $r_t=0.01$~m, a wall inflection angle $\alpha_i=9^{\circ}$, 
a base area ratio $\epsilon_B=r_b^2/r_t^2=11.3$,
an extension area ratio $\epsilon_E=r_e^2/r_t^2=27.1$, with the extension displaying a constant 
wall-pressure profile (CP). 
The nozzle pressure ratio is fixed at the value 45.7, since at this NPR
the nozzle flow experiences a symmetric shock separation located exactly at the
inflection point. The other parameters of the simulation were chosen to reproduce
the operating conditions of the experimental campaign. In particular, the total
temperature $T_0$ and the static ambient temperature $T_a$ have been set equal to
300 K. The nozzle Reynolds number is

\begin{equation}
        Re = \frac{\rho_0 a_0 r_t}{\mu_0} = \frac{\sqrt{\gamma}}
        {\mu_0}\frac{p_0 r_t}{\sqrt{R_{air} T_0}}=1.03 \cdot 10^7,
\end{equation}

where $\rho_0$ is the density, $a_0$ the speed of sound and $\mu_0 = \mu (T_0)$
is the molecular viscosity taken at the stagnation-chamber condition.
The three-dimensional computational domain includes the external ambient (see
Fig.~\ref{fig:mesh}) that extends up to 150 $r_t$ in the longitudinal direction and
80 $r_t$ in the radial one from the symmetry axis.
The boundary conditions are imposed as follows:
total temperature, total pressure and flow direction are enforced at the nozzle
inflow, while on the outflow boundary at the end of the external domain a non-reflecting boundary
condition is implemented.
A back-pressure equal to the ambient pressure $p_a$ is imposed at the other boundaries.
The no-slip adiabatic condition is prescribed for the nozzle walls.
A grid sensitivity analysis was carried out to determine the proper
mesh resolution. The study was performed by means of steady-state axi-symmetric
RANS simulations and the RANS solution was also used to initialise the three-dimensional DDES simulation.
The development of turbulent structures and the passage from modelled to resolved turbulence was
stimulated by adding random perturbations to the streamwise velocity field at the
initial time. Those perturbations have a maximum magnitude of $3\%$ of the inflow velocity.
The computational domain is composed by 136 structured blocks in the $x-y$ plane, each block discretized with
$22\times256$ cells. The 3D mesh includes 256 cells in the azimuthal direction,
for a total number of approximately 196 million cells.
The computation was run with a time step $\Delta t = 4.65 \cdot 10^{-8}$s and a
relatively long time span was simulated $T = 0.0172$ s, which
guarantees coverage of frequencies down to at least $f_{min} \approx 58 $ Hz.
A total of 80 full three-dimensional fields have been collected at time intervals of
$2.15 \cdot 10^{-4}$ s for post-processing purposes. Furthermore, samples of the
pressure field at the wall and in an azimuthal plane have been recorded at shorter
time intervals of $2.34 \cdot 10^{-6}$ s to guarantee sufficient resolution for the
frequency analysis.

In the following, the results for the dual bell are compared with those recently obtained
for a truncated ideal contour (TIC) nozzle with free-shock separation (FSS) operating at NPR = 30.35,
experimentally tested at the University of Texas at Austin and
simulated with the same methodology here exposed~\citep{martelli2019flow}.
Basic geometric properties of the TIC nozzle are a throat radius of
$r_t = 0.019.0$ m, an exit radius of $r_e = 0.117$ m and a throat-to-exit length of $L = 0.351 $ m.
A summary of the main geometrical parameters of both the dual bell and TIC nozzle is reported
in Table~\ref{tab:DBTIC}. An extensive analysis on the flow unsteadiness and wall-pressure signature
for the TIC nozzle can be found in our previous paper~\citep{martelli2019flow}.
Here, novel results for that geometry are presented concerning the analysis of the 
aerodynamic side loads, compared in Sec.~\ref{sec:sl} with those generated in the dual-bell nozzle.
\begin{table}
\centering
\caption{Dual-Bell and TIC nozzles geometrical parameters.}
\label{tab:DBTIC}
\begin{tabular}{lccc}
\hline\noalign{\smallskip}
 &  & DB & TIC \\
\noalign{\smallskip}\hline\noalign{\smallskip}
Throat radius & $r_t$         & 10 mm & 19 mm \\
Area ratio    & $\epsilon_E$  & 27.1  & 38    \\
Total length  & $L_{tot}/r_t$ & 14.24 & 18.44 \\
\noalign{\smallskip}\hline
\end{tabular}
\end{table}

\section{Results}

\subsection{Flowfield organisation}
\label{sec:flow_org}

\begin{figure*}
  \centering
  \includegraphics[width = 0.30\textwidth,angle=270,clip]{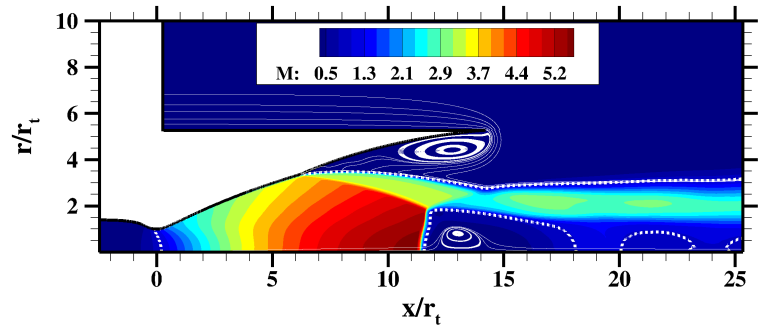} \vskip 1em
  \caption{Contours of the averaged Mach number field from DDES. The white dashed line denotes the sonic level.}
  \label{fig:mean_mach}
 \end{figure*}

 \begin{figure*}
  \centering
(a)\includegraphics[width = 0.35\textwidth,angle=270,clip]{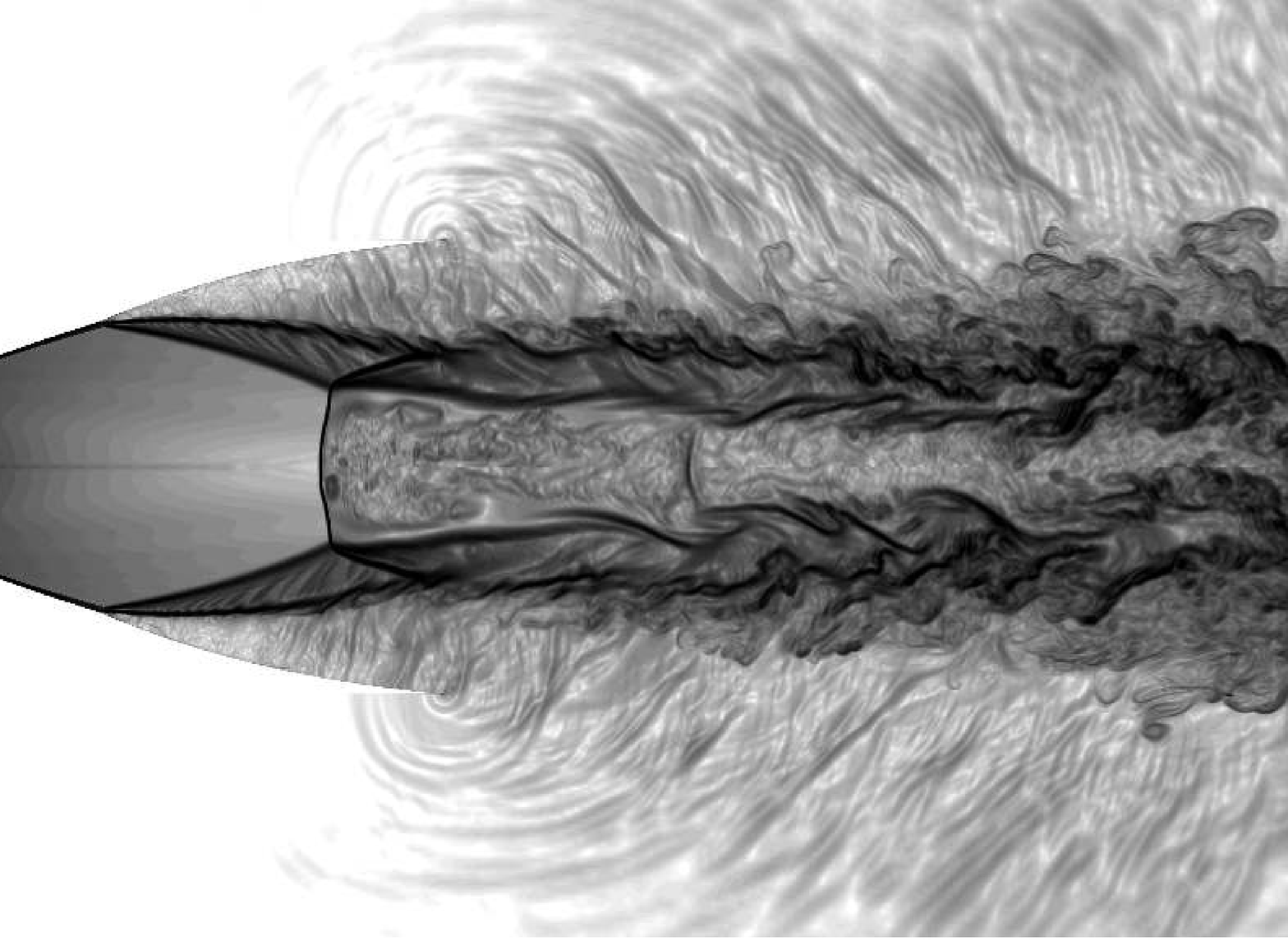} \vskip 1em
(b)\includegraphics[width = 0.25\textwidth,angle=270,clip]{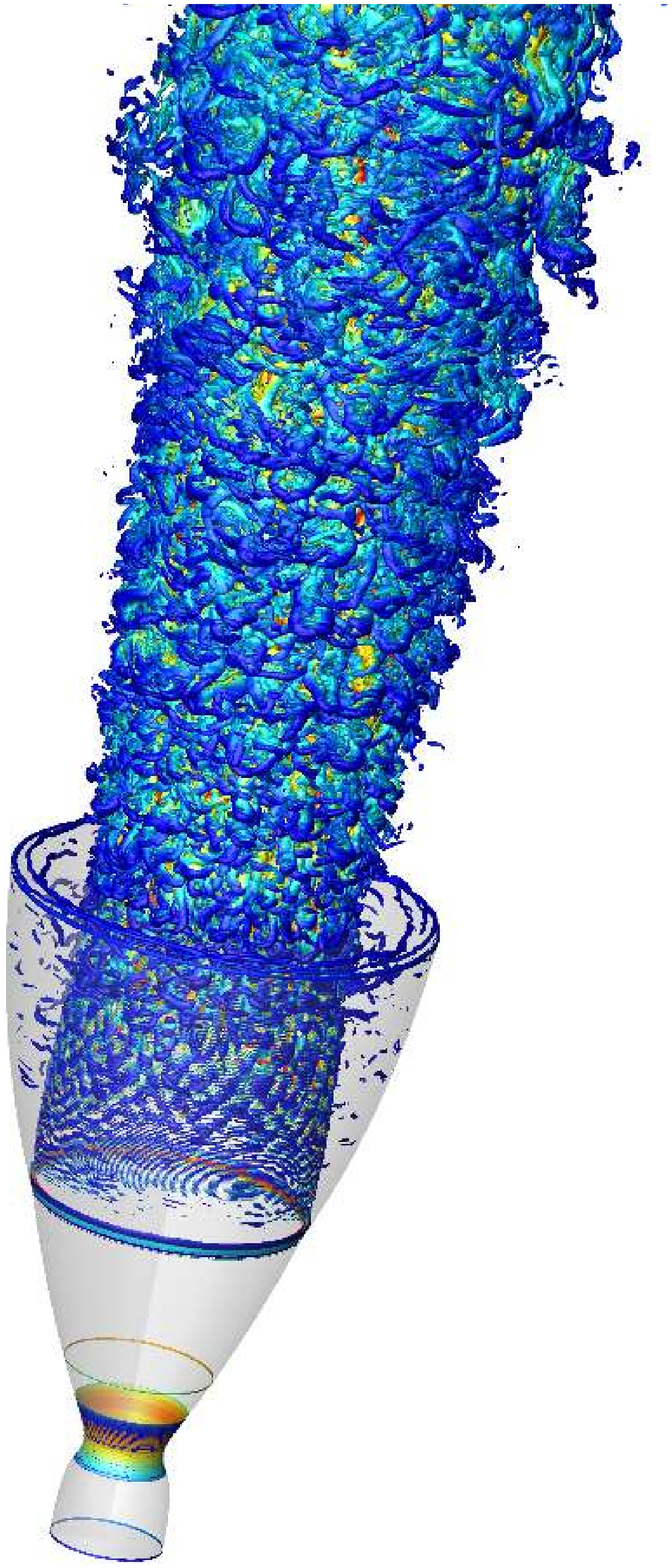} \vskip 1em
  \caption{Visualisation of (a) the instantaneous density-gradient magnitude (numerical Schlieren) in an x-y plane
  and of (b) turbulent structures through an iso-surface of the Q criterion, coloured with the local value
of the longitudinal velocity.}
  \label{fig:ddes45.7}
 \end{figure*}

The main features of the flow pattern inside the dual-bell nozzle are reported
in Fig.~\ref{fig:mean_mach}, which shows a longitudinal $x-r$ plane with
the iso-contours of the mean Mach-number field,
obtained by averaging in time and in the azimuthal direction.
At the selected NPR, the flow is separated and anchored at the inflection point,
generating a free shear layer. A shock system arises comprising the conical separation-shock,
a Mach-disk reflection and a second conical shock, which re-directs the shear layer
in a direction almost parallel to the nozzle axis.
The dual-bell extension is characterized by a considerable subsonic turbulent recirculating 
region. 
The flow downstream of the Mach disk is initially subsonic, 
then it experiences an expansion process across a fluid-dynamic throat, 
that is visible from the sonic line, and again accelerates to a supersonic velocity.
Then a new shock appears to balance the pressure of the flow to the ambient level.
It is also interesting to observe the presence of a recirculating region near 
the nozzle axis downstream the Mach disk. This region is fed by the vorticity produced by 
the shock curvature, which causes an entropy gradient in the radial direction and hence vorticity according
to the Crocco's theorem.

The unsteadiness of the flow is highlighted by Fig.~\ref{fig:ddes45.7} (a), where
contours of the density-gradient magnitude are displayed in a longitudinal $x-r$ plane.
The developing turbulent structures of the two co-annular supersonic shear layers 
are well visible. The external shear layer originates from the jet detachment from 
the wall and the internal one originates from the triple point of the Mach reflection.
The two layers merge downstream in the external ambient, where the Mach waves irradiated 
by the vortices are well visible. It is also possible to observe fine turbulent 
structures emitted by the Mach disk, coherently with the observation reported above on the 
entropy gradient generation due to the shock bending.
The Q-criterion is generally used to identify the tube-like structures
from a qualitative point of view~\citep{hunt88}. 
In this work a modified definition of the Q-criterion
has been adopted to include the effects of compressibility~\citep{pirozzoli2008}.
An isosurface of the Q-criterion is shown in Fig.~\ref{fig:ddes45.7} (b), 
coloured according to the value of the streamwise velocity component. It shows that 
the Kelvin-Helmotz instability of the  initial part of the shear layer 
is not characterised by the coherent  toroidal vortices which can be found 
in incompressible flows. Instead, we observe that oblique modes dominate the initial part of the
shear layer, then leading to the generation of small-scale three-dimensional structures,
well resolved by the present DDES approach.
The differences in the initial shear layer development observed with respect to the
typical pattern of low-speed flows can be attributed to the large local convective Mach number ($M_c \approx 1$) 
at the beginning of the detached shear layer, which changes the shear-layer
instability process~\citep{sandham_reynolds_1991}.
A similar behaviour was observed by~\citet{martelli2019flow} in a conventional TIC nozzle with flow separation  
and by~\citet{deck2007} in a supersonic cylindrical base flow.

\subsection{Analysis of the wall-pressure signature}
\label{sec:wallp}


 \begin{figure*}
  \centering
  \psfrag{X}[c][c][1.2]{$x/r_t$}
  \psfrag{Y}[c][c][1.2]{$\overline{p_{w}}/p_{a}$}
(a)\includegraphics[width = 0.3\textwidth,angle=270,clip]{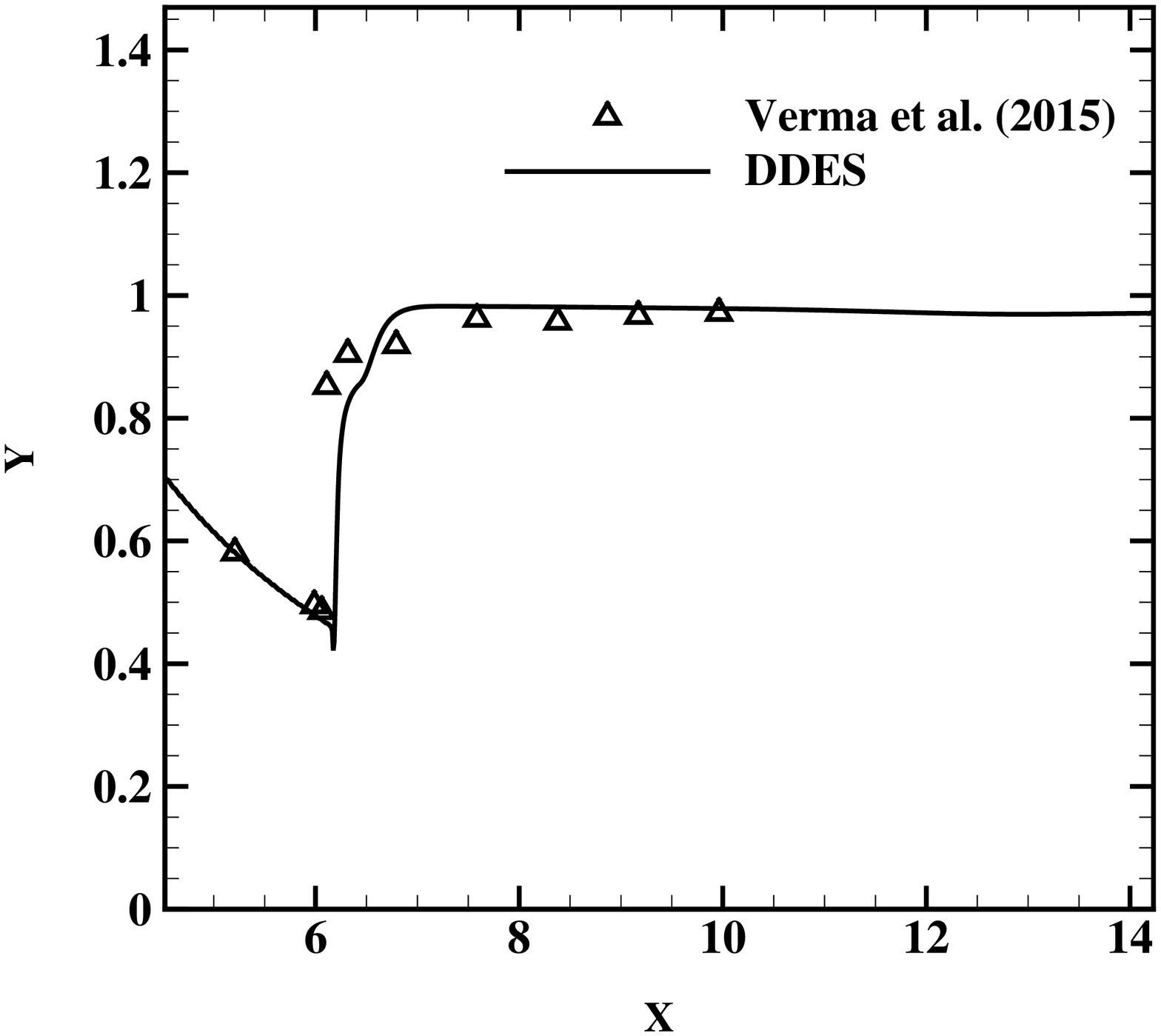} \hskip 1em
  \psfrag{X}[c][c][1.2]{$x/r_t$}
  \psfrag{Y}[c][c][1.2]{$\sigma_{w}/p_{w}$}
(b)\includegraphics[width = 0.3\textwidth,angle=270,clip]{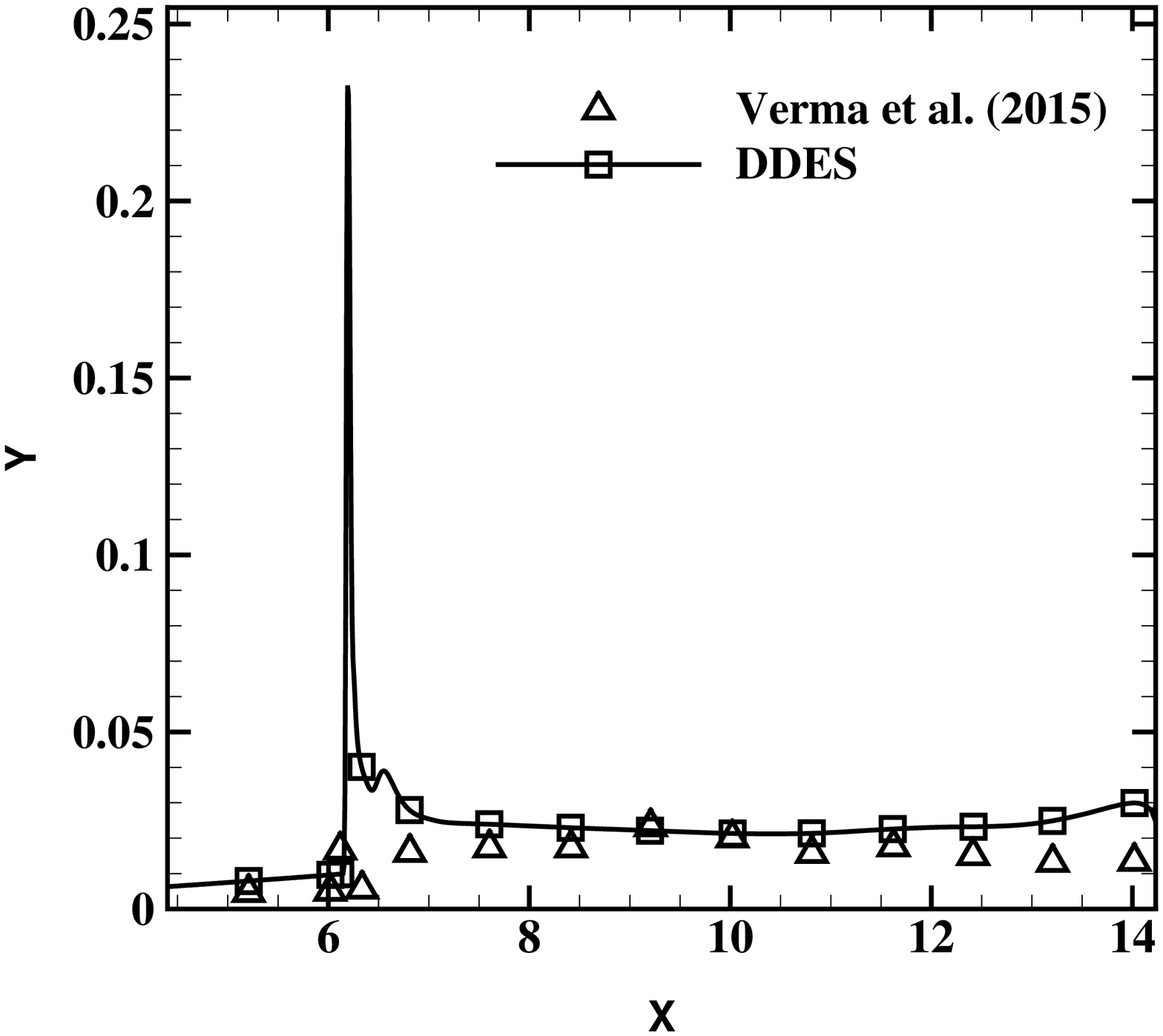} \vskip 1em
	 \caption{Comparison of the distribution of the normalized mean wall pressure (a) and 
	          standard deviation of wall-pressure fluctuations (b) with experimental data.}
  \label{fig:pw_verma}
 \end{figure*}

The spatial evolution of the mean wall pressure ($\overline{p}_w/p_a$),
averaged in time and in the azimuthal direction,
is presented in Fig.~\ref{fig:pw_verma}(a), compared with the reference experimental
data~\citep{verma2015}.
The mean wall pressure decreases until the separation point, which is located
at the inflection point, then the oblique shock causes a sudden increase up
to a plateau value close to the ambient pressure level. The numerical data are in
good agreement with the experimental ones. In particular, the right values of 
wall pressure in the turbulent recirculating zone indicates that
the LES-branch simulation is able to correctly capture the momentum exchange between the main jet
and the separated flow, dominated by the high convective Mach number of the shear
layer~\citep{pantano02}.

The standard deviation of the wall-pressure signal ($\sigma_{w}/p_{w}$) is shown in
Fig.~\ref{fig:pw_verma}(b) and compared with the corresponding experimental values. The
distribution of $\sigma_{w}/p_{w}$ along the  longitudinal axis is typical of shock-wave/turbulent boundary
layer interactions~\citep{Dolling1985}, characterised by a dominant sharp peak at
the separation-shock location and followed by a lower level in the zone where the
turbulent shear layer develops.
The numerical standard deviation agrees with the experimental behaviour, especially in the turbulent 
recirculating zone. It seems that the sharp peak predicted by DDES at the shock location
is not captured by the experimental measurements, probably due to the probe spacing.
Indeed, the flow separation is fixed at the inflection point, 
therefore the pressure fluctuations are confined in a very narrow spatial range.
\begin{figure}
  \centering
  \psfrag{X}[c][c][1.2]{$x/r_t$}
  \psfrag{Y}[c][c][1.2]{$\sigma_{w}/p_{w}$}
  \includegraphics[width = 0.3\textwidth,angle=270,clip]{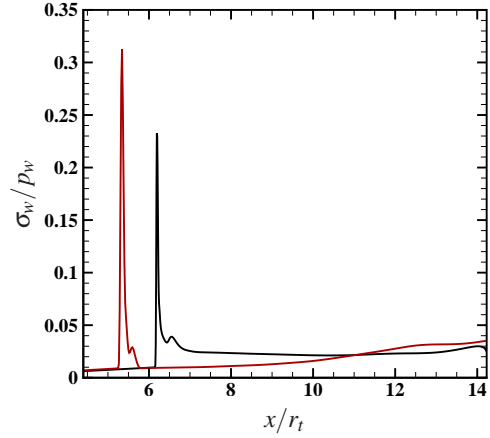} \vskip 2em
  \caption{Comparison between dual-bell and TIC nozzles of the standard deviation of wall-pressure 
	fluctuations. The black solid line refers to the dual-bell nozzle while the red solid one to the TIC.}
  \label{fig:rms_tic}
\end{figure}
The dual-bell standard deviation of the wall-pressure fluctuations  
is also compared with that of the TIC nozzle in Fig.~\ref{fig:rms_tic}.
The standard deviations have the same global behaviour, 
with the TIC characterised by a higher value in correspondence of the peak and
by a more marked drop downstream of the shock location. 
Then the energy level of the pressure 
fluctuation gradually increases during the development of the shear layer. 
The dual-bell nozzle instead has a more flat trend downstream of the peak. 
 \begin{figure*}
  \centering
(a)\includegraphics[width = 0.40\textwidth,angle=270,clip]{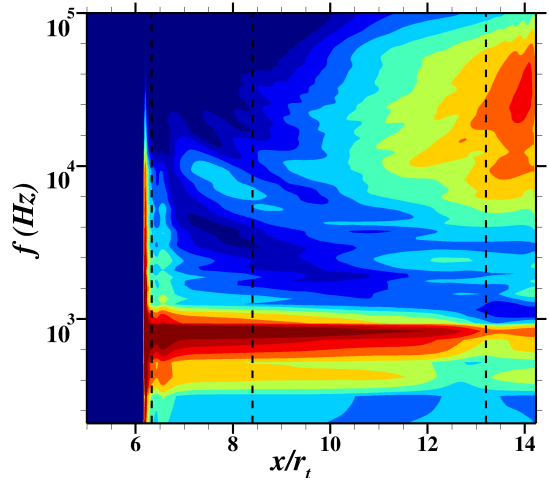} 
(b)\includegraphics[width = 0.40\textwidth,angle=270,clip]{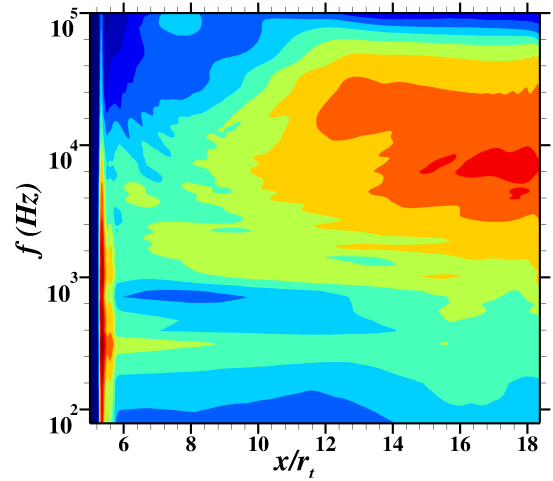} \hskip 1em
	 \caption{Contours of premultiplied power spectral densities $G(f)\cdot f/\sigma^2$ 
	 of the wall-pressure as a function of the streamwise location and frequency: 
	 (a) Dual-Bell nozzle and (b) TIC nozzle. Eleven contour levels are shown in 
	 exponential scale between 5$\cdot$10$^{-4}$ and 0.1. The dashed black lines in (a) denote the 
	 probes chosen for the comparison with experimental data.}
  \label{fig:mapspec}
 \end{figure*}
\begin{figure}
  \centering
  \psfrag{X}[c][c][1.2]{$f\,(Hz)$}
  \psfrag{Y}[c][c][1.2]{$f\,G(f)/\sigma^2$}
  \includegraphics[width = 0.30\textwidth,angle=270,clip]{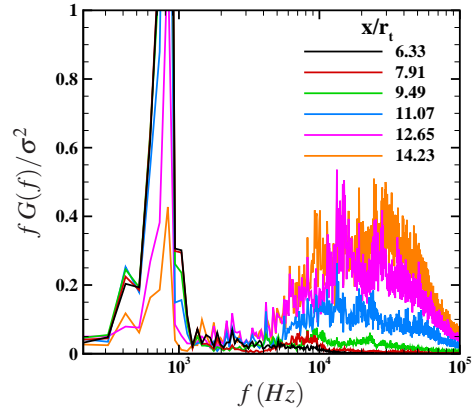} \vskip 1em
  \caption{Axial evolution of the normalized premultiplied power spectral densities, $G(f)\cdot f/\sigma^2$ 
	   of the wall-pressure signals.}
  \label{fig:specx_axial}
 \end{figure}
%
%
\begin{table}
\centering
\caption{Comparison of the shock motion peak frequencies with the experimental data.}
\label{tab:t nepeak_vermacomp}
\begin{tabular}{ccc}
\hline\noalign{\smallskip}
$x/r_t$ & DDES & Exp.~\citep{verma2015} \\
\noalign{\smallskip}\hline\noalign{\smallskip}
6.33  & 833 Hz & 775 Hz \\
8.40  & 833 Hz & 764 Hz \\
13.20 & 833 Hz & 787 Hz \\
\noalign{\smallskip}\hline
\end{tabular}
\end{table}

To provide a global picture of the pressure energy distribution
along the nozzle wall and in the frequency domain,
contours of the premultiplied wall-pressure spectra $G(f) \, f/\sigma^2$ are
reported in Fig.~\ref{fig:mapspec}(a) with respect to the longitudinal
coordinate $x/r_t$ and frequency $f$. 
From the spectral map, it is possible to observe the presence of two regions characterised by high fluctuation energy.
The first zone of interest is located near the separation point ($x/r_t = 6.33$) and
is characterised by a broad bump in the low-medium frequency range and
very narrow in the longitudinal direction. It is clearly
associated to the signature of the unsteady shock motion.
The peak of this bump is at a frequency of $\sim$830 Hz and its footprint is well visible in
the spatial direction until the end of the nozzle. 
According to~\citet{Baars2012} and~\citet{martelli2019flow} the low-frequency peak 
can be attributed to an acoustic resonance, which can be described by 
the one-quarter acoustic standing wave model~\citep{wong2005} in open-ended pipes. 
The resonance frequency $f_{ac}$ can be expressed as follows,
\begin{equation}
	f_{ac}=\frac{a_\infty \, \left(1-M^{2}_{N.E.}\right)}{4(L+\varepsilon)}
        \label{eq:fac}
\end{equation}
where $a_\infty=345 m/s$ is the ambient speed of sound, $M_{N.E.}=0.023$ is the Mach
number in the separated region at the nozzle exit,
$L=l_e=0.0804 \, m$ is the distance between the inflection point and the nozzle lip.  
This distance should correspond to the pipe length in the 1D acoustic model
 and $\varepsilon$ is an opened-end empirical correction parameter~\citep{proschanka2012}. 
According to ~\citet{proschanka2012} this parameter can vary between $0.5L$ and $0$ giving 
a range of frequency between $715$ Hz and $1072$ Hz, which include 
the numerical peak frequency of $833$ Hz.  
It should also be considered that the presence of a large turbulent recirculating region
increases the difference of the present situation with the acoustic pipe 1D model.

The second zone characterised by high levels of pressure fluctuations is located in
the high-frequency range (between $f = 10^4$ and $10^5$ Hz) and is associated with the
development of the separated shear layer, from which pressure disturbances are
emitted through the vortices that are convected downstream at high speed.

For a better quantification of the evolution of the wall-pressure frequency content
along the longitudinal axis, Fig.~\ref{fig:specx_axial} 
shows the premultiplied spectra at six axial equispaced probes.
The peaks associated to the shock movement have very high amplitudes
(for visualisation purposes the values are cut at 1) at $\approx$ 833 Hz and they
persists for almost  all the stations considered.
Starting from the first probe the energy content 
of the shock peak tends to reduce along the x-axis
whereas the energy contribution of the separated shear layer rises 
in amplitude at higher frequencies.
Table~\ref{tab:t nepeak_vermacomp} reports the peak frequencies of the spectra 
(attributed to the acoustic resonance) at the three different axial stations corresponding 
to the location of the experimental probes (black dashed lines in Fig.\ref{fig:mapspec}). 
There is a rather good agreement between the numerical results (833 Hz) and the experiments 
(between 775 Hz and 787 Hz), and in addition~\citet{verma2015} shows that the acoustic tone 
persists all along the second bell also in the experiment.

Fig.~\ref{fig:mapspec}(b) shows the contour map of the wall-pressure spectra for the TIC 
nozzle, as also reported in~\citet{martelli2019flow}. Qualitatively the two maps are very 
similar. The main quantitative difference is that the footprint of the shock oscillation decreases 
much rapidly in the longitudinal direction with respect to the dual-bell case.
Furthermore, the TIC wall-pressure signature presents some
non-negligible energy in the intermediate frequency range (around 1 kHz), whereas the 
dual-bell nozzle show a very small energy content 
in the medium frequency range (around $\approx$ 2300 Hz).
As previously speculated by~\citet{Jaunet2017} and then
confirmed by~\citet{martelli2019flow}, this intermediate-frequency peak
can be attributed to a screech-like mechanism with a helical-type mode occurring inside
the nozzle. This aspect is deeply discussed in the following section.


\subsection{Azimuthal decomposition of the pressure field}
\label{sec:azimuthal}

\begin{figure*}
  \centering
         \psfrag{M}[c][c][1.]{$0^{th}\,mode$}
\includegraphics[width = 0.10\textwidth,angle=270,clip]{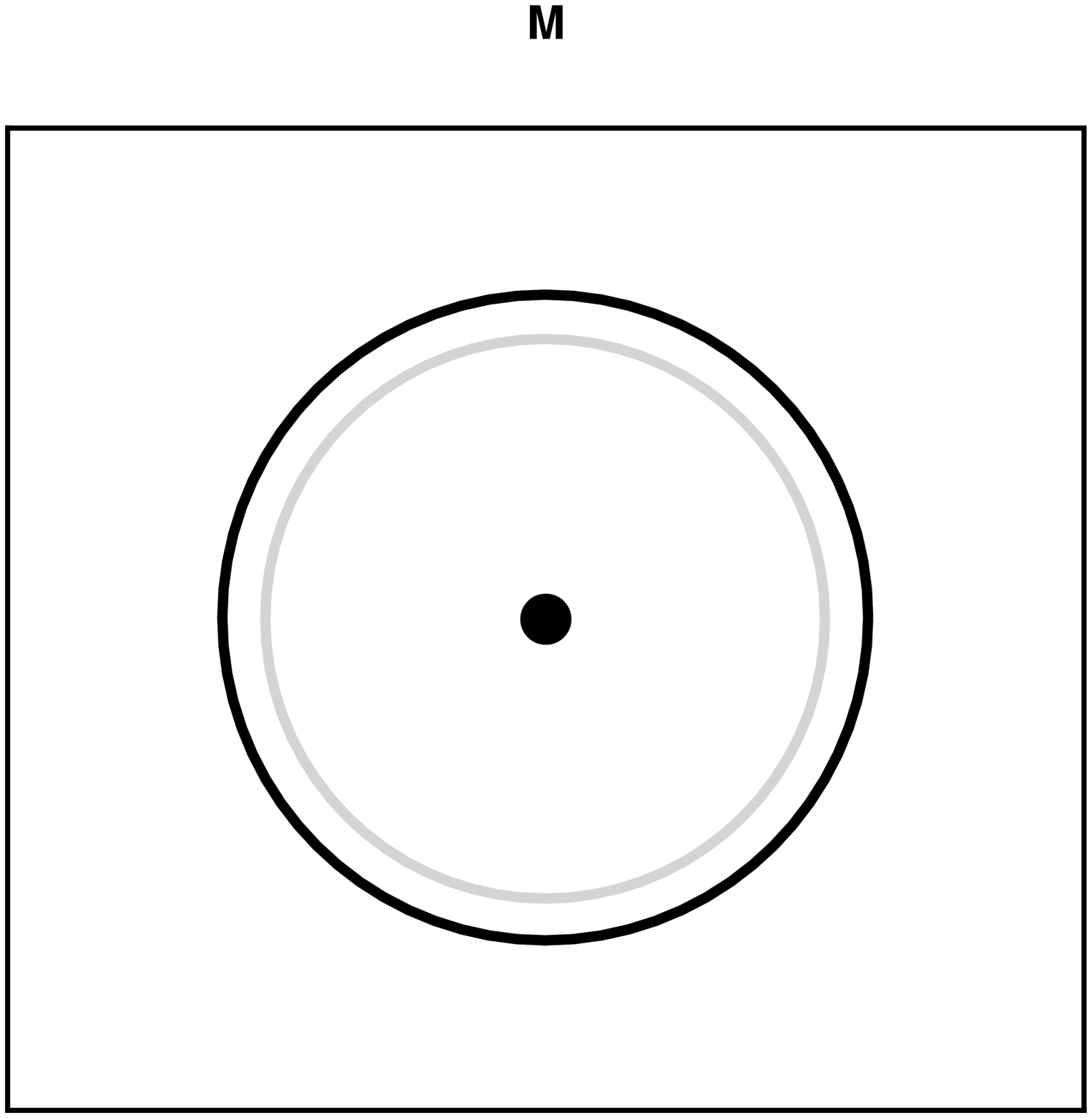} 
         \psfrag{M}[c][c][1.]{$1^{st}\,mode$}
\includegraphics[width = 0.10\textwidth,angle=270,clip]{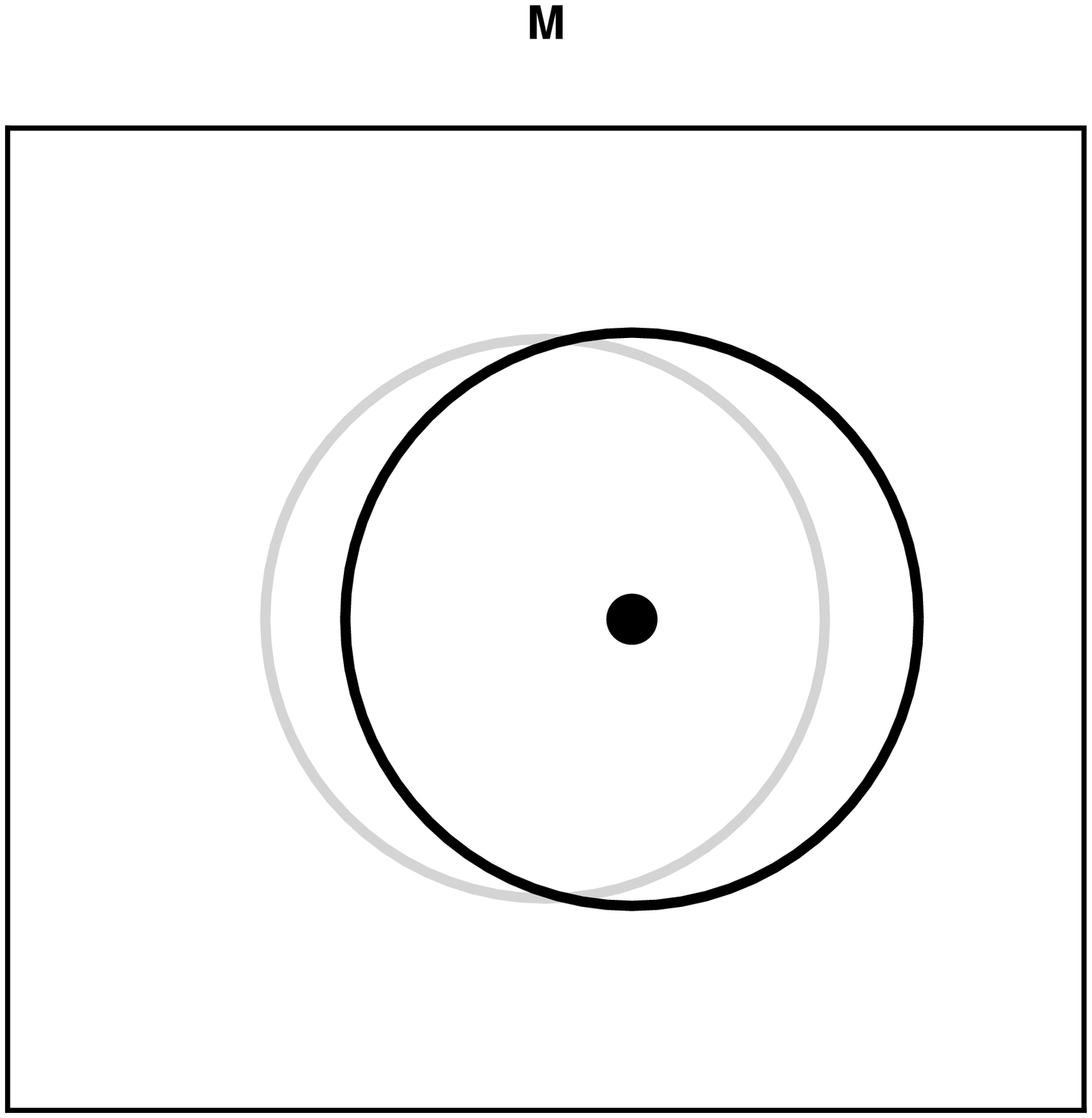} 
         \psfrag{M}[c][c][1.]{$2^{nd}\,mode$}
\includegraphics[width = 0.10\textwidth,angle=270,clip]{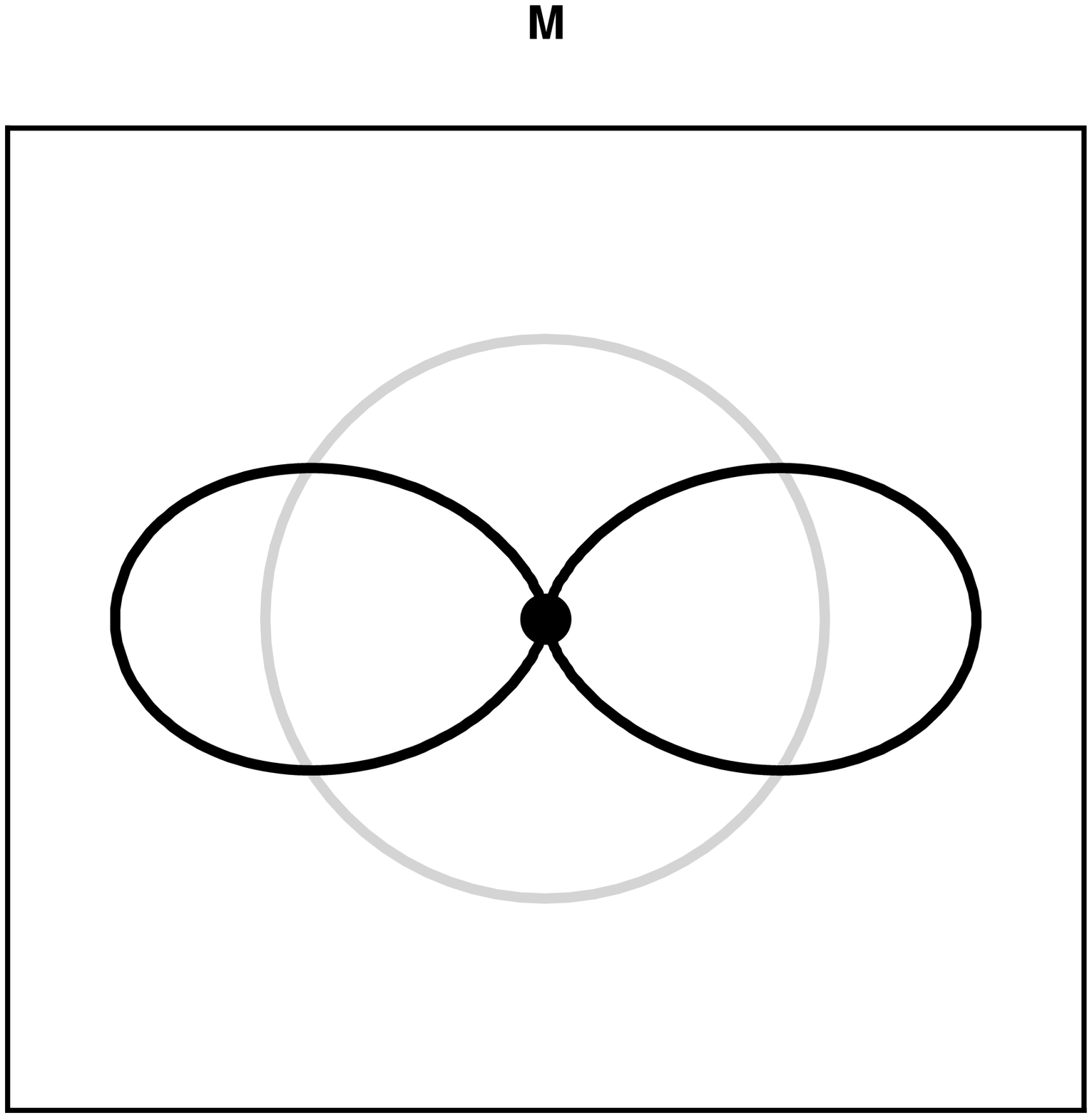} 
         \psfrag{M}[c][c][1.]{$3^{rd}\,mode$}
\includegraphics[width = 0.10\textwidth,angle=270,clip]{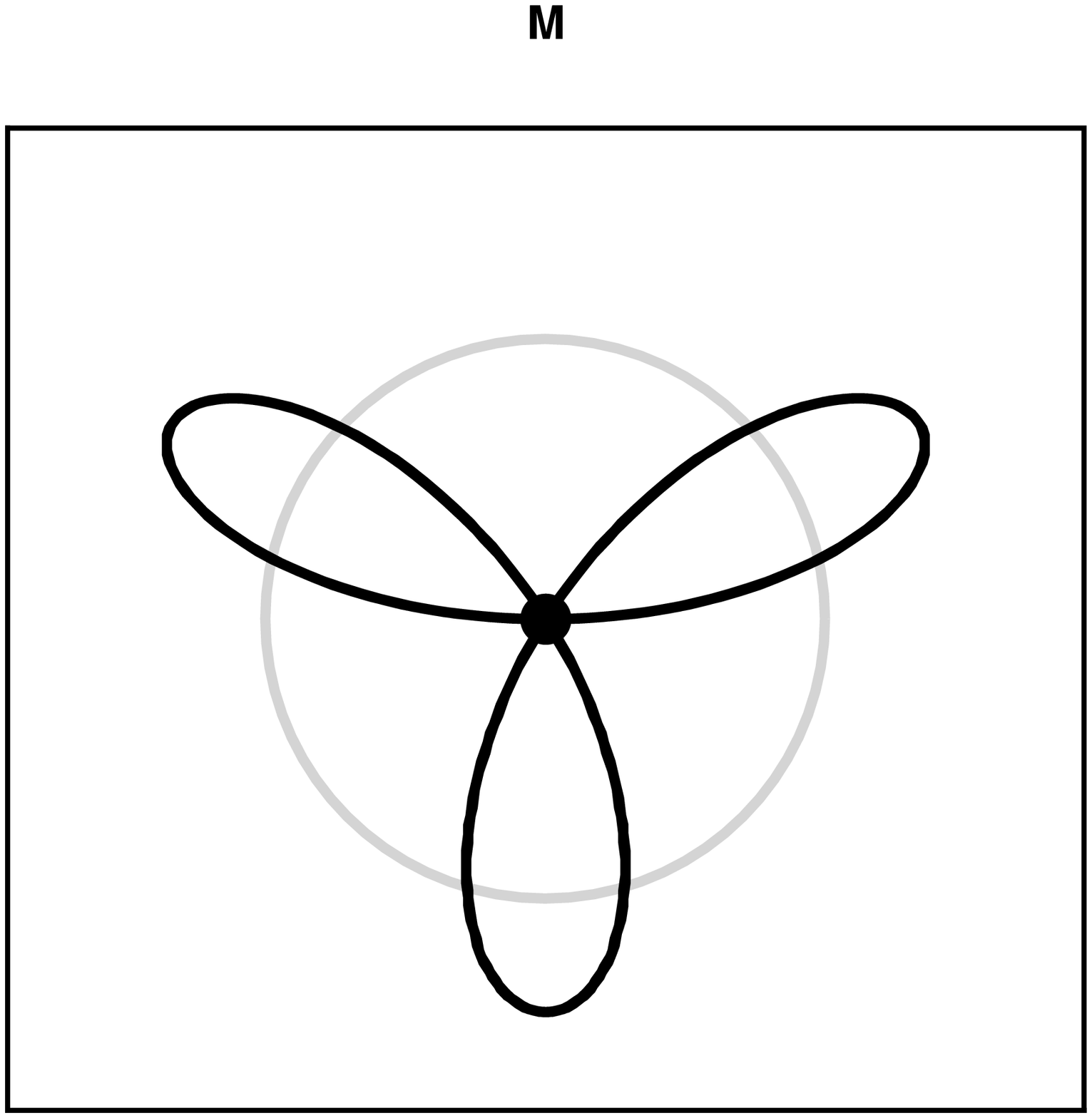} 
         \psfrag{M}[c][c][1.]{$4^{th}\,mode$}
\includegraphics[width = 0.10\textwidth,angle=270,clip]{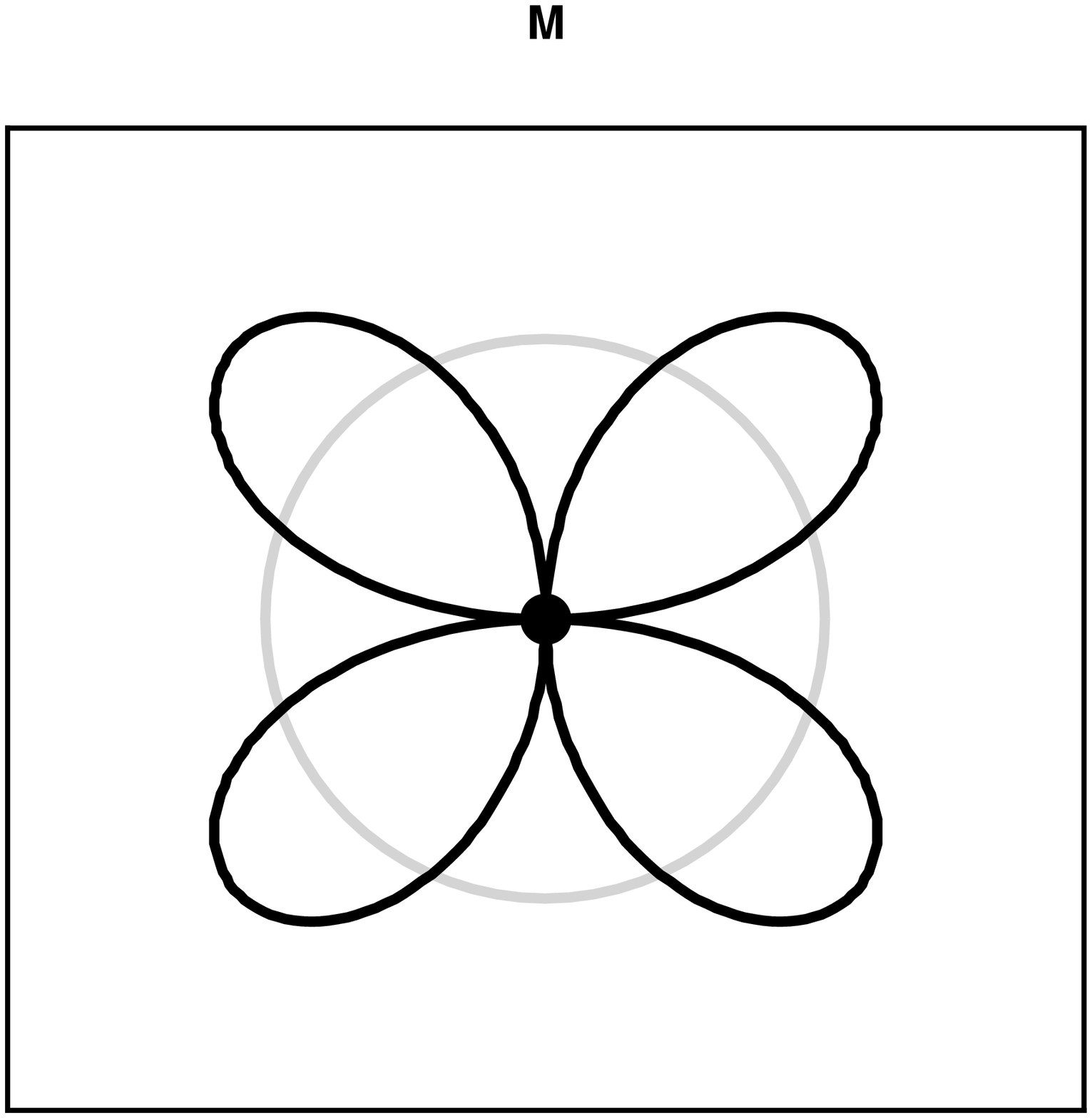} 
        \caption{Physical interpretation of the Fourier azimuthal decomposition. Solid black lines
                 refer to the real ($m>0$) part of the Fourier coefficients, solid grey lines
                 refer to the unit circle.}
  \label{fig:modes_sketch}
\end{figure*}

%

\begin{figure}
  \centering
  \psfrag{Y}[c][c][1.2]{$\alpha(x,m)$}
  \psfrag{X}[c][c][1.2]{$m$}
(a)\includegraphics[width = 0.3\textwidth,angle=270,clip]{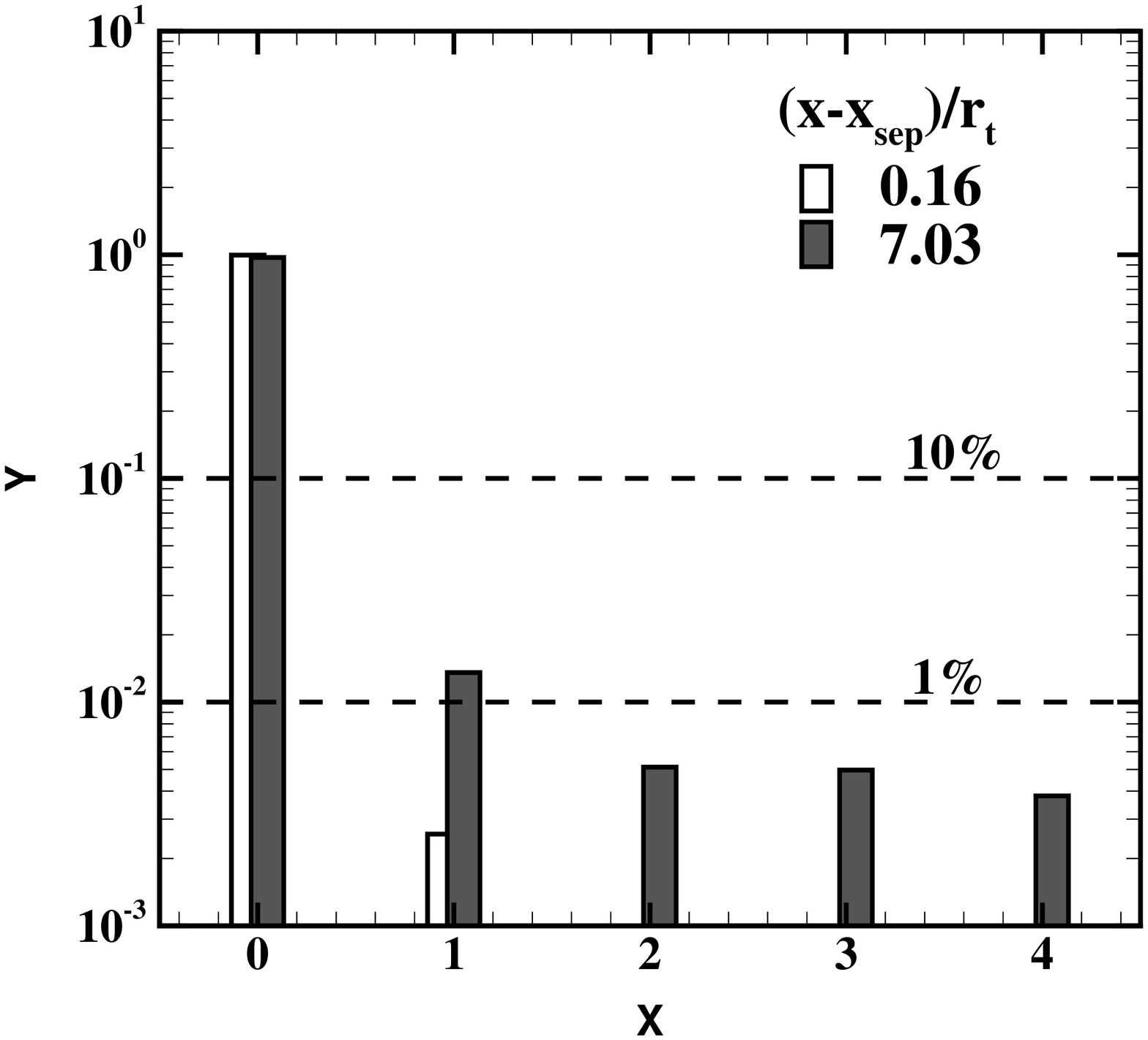} \hskip 1em
(b)\includegraphics[width = 0.3\textwidth,angle=270,clip]{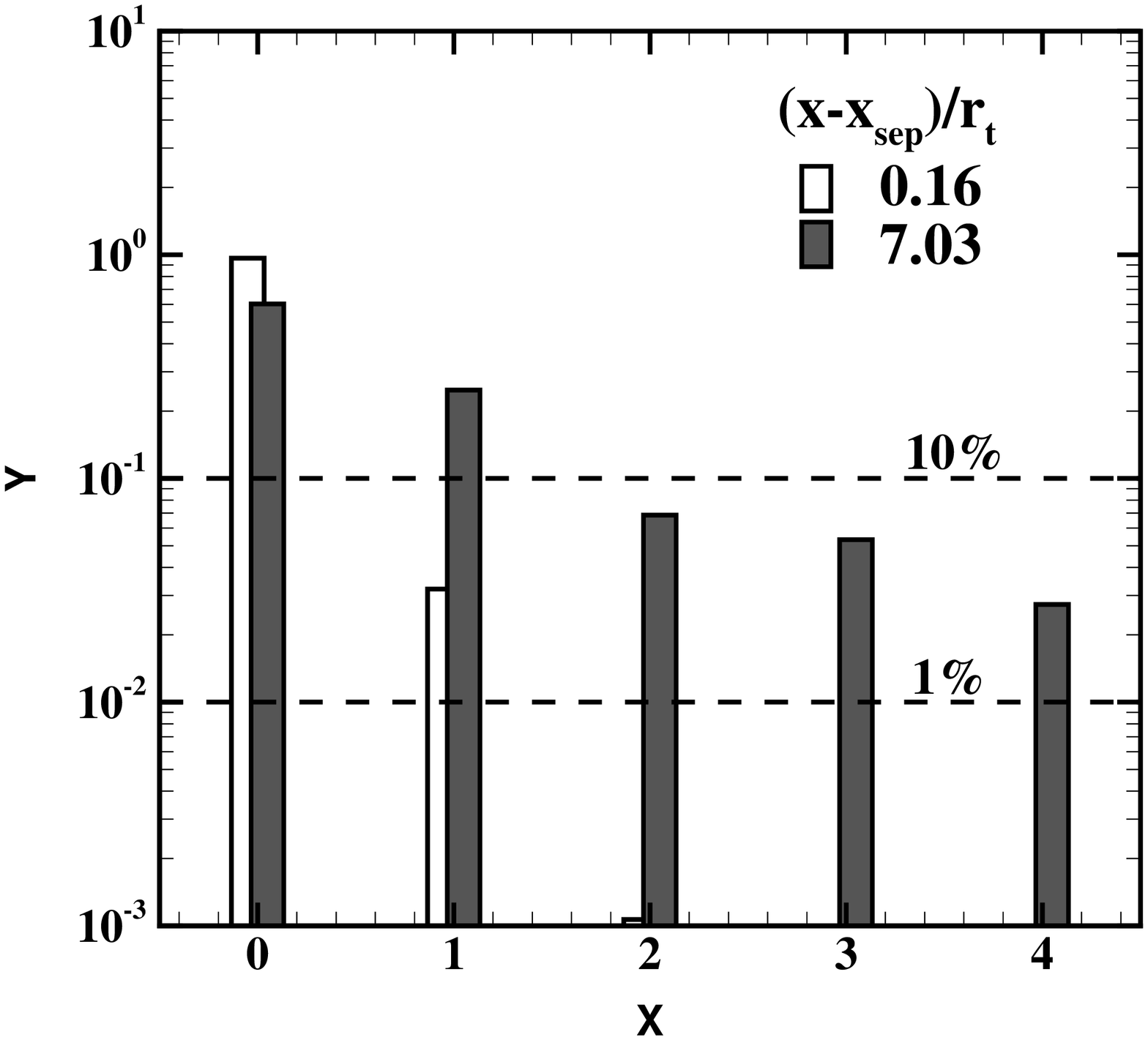} \hskip 1em
  \caption{Normalized eigenspectra of the Fourier modes as fraction of resolved energy per mode at two
           different axial stations: (a) dual-bell nozzle and (b) TIC nozzle.}
  \label{fig:mode_histo}
 \end{figure}
\begin{figure*}
  \centering
         \psfrag{X}[c][c][1.2]{$St$}
         \psfrag{Y}[c][c][1.2]{$f\phi_{pp}\, (psi^2)$}
(a)\includegraphics[width = 0.3\textwidth,angle=270,clip]{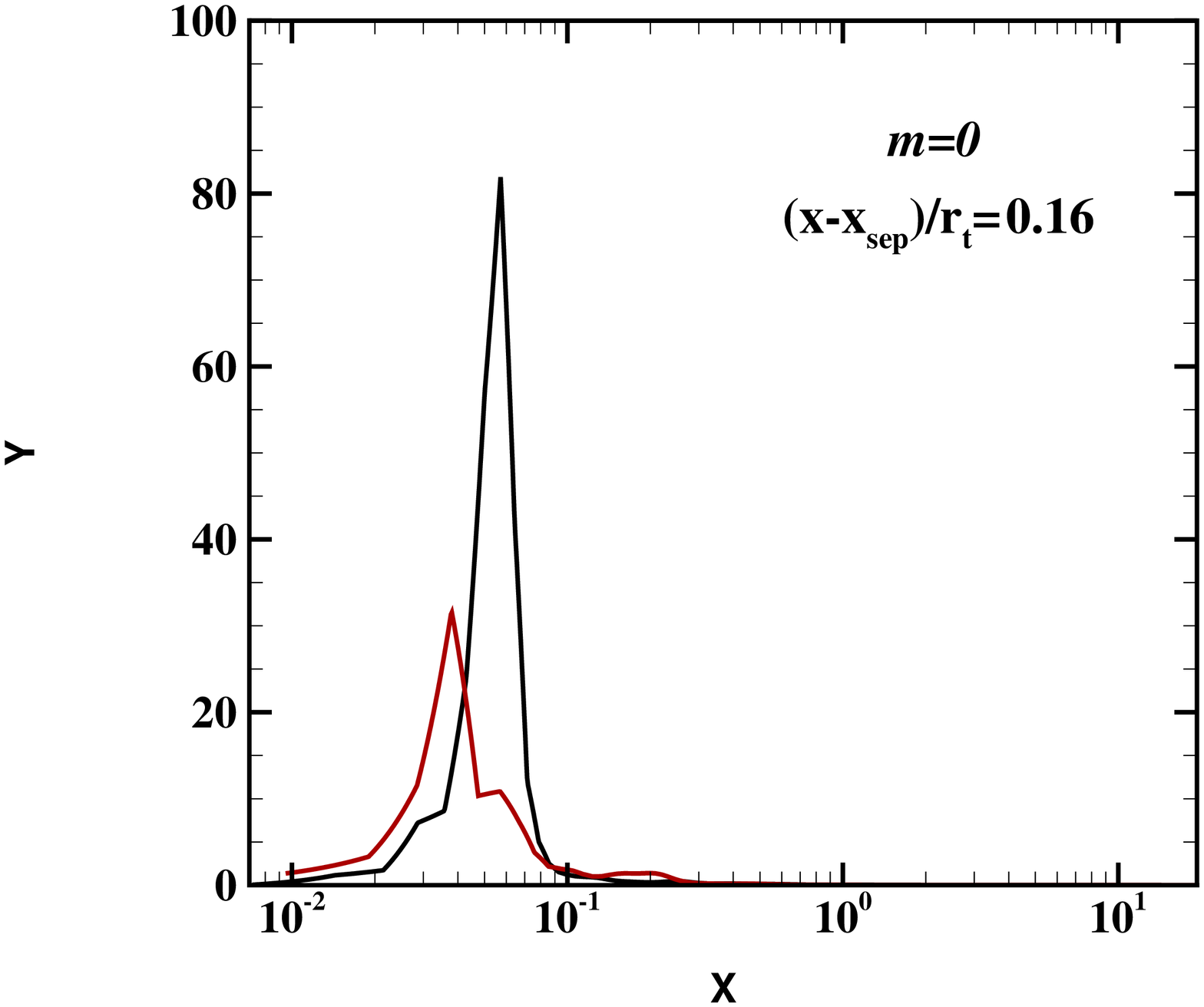} \hskip 1em
(b)\includegraphics[width = 0.3\textwidth,angle=270,clip]{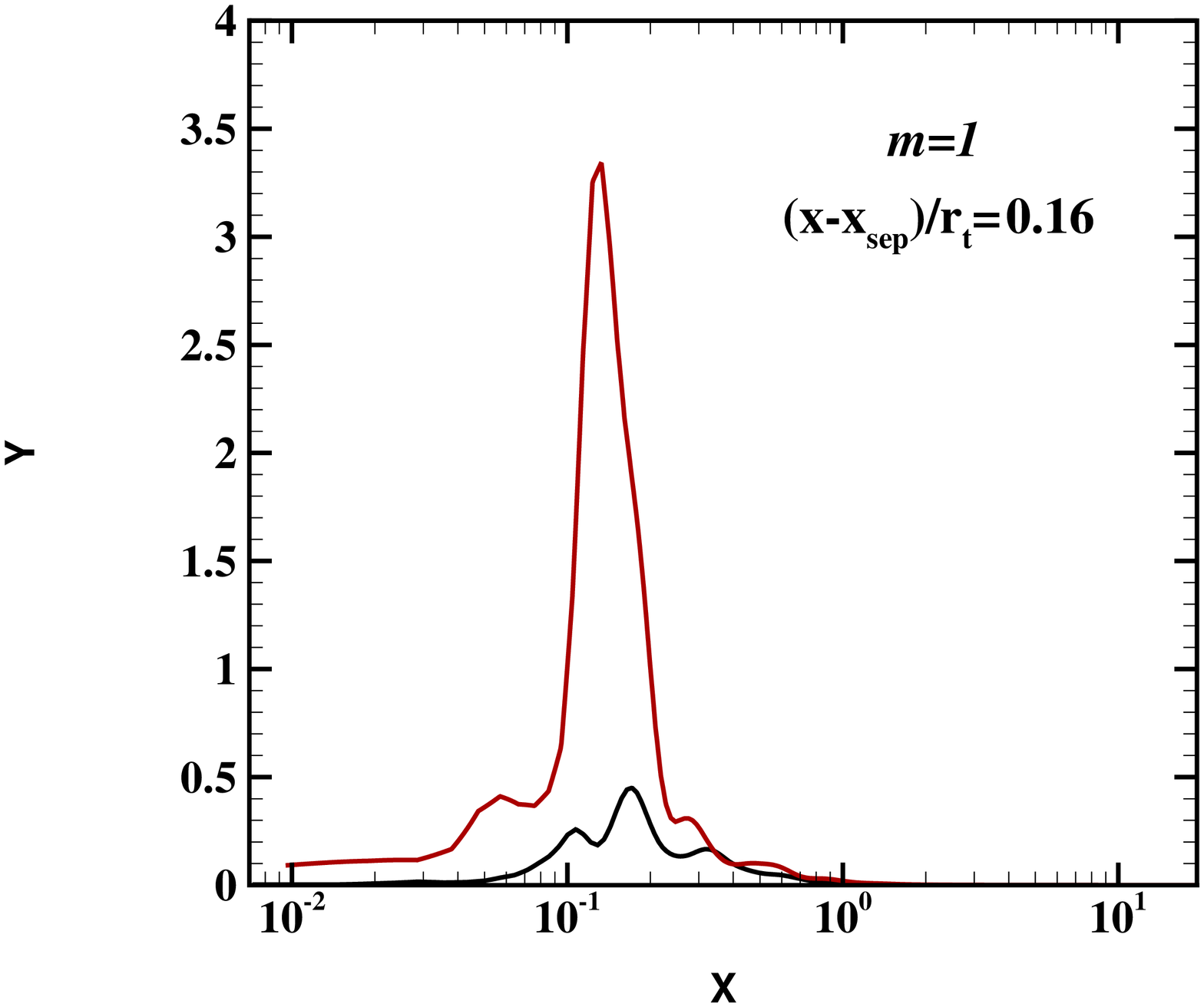} \vskip 1em
(c)\includegraphics[width = 0.3\textwidth,angle=270,clip]{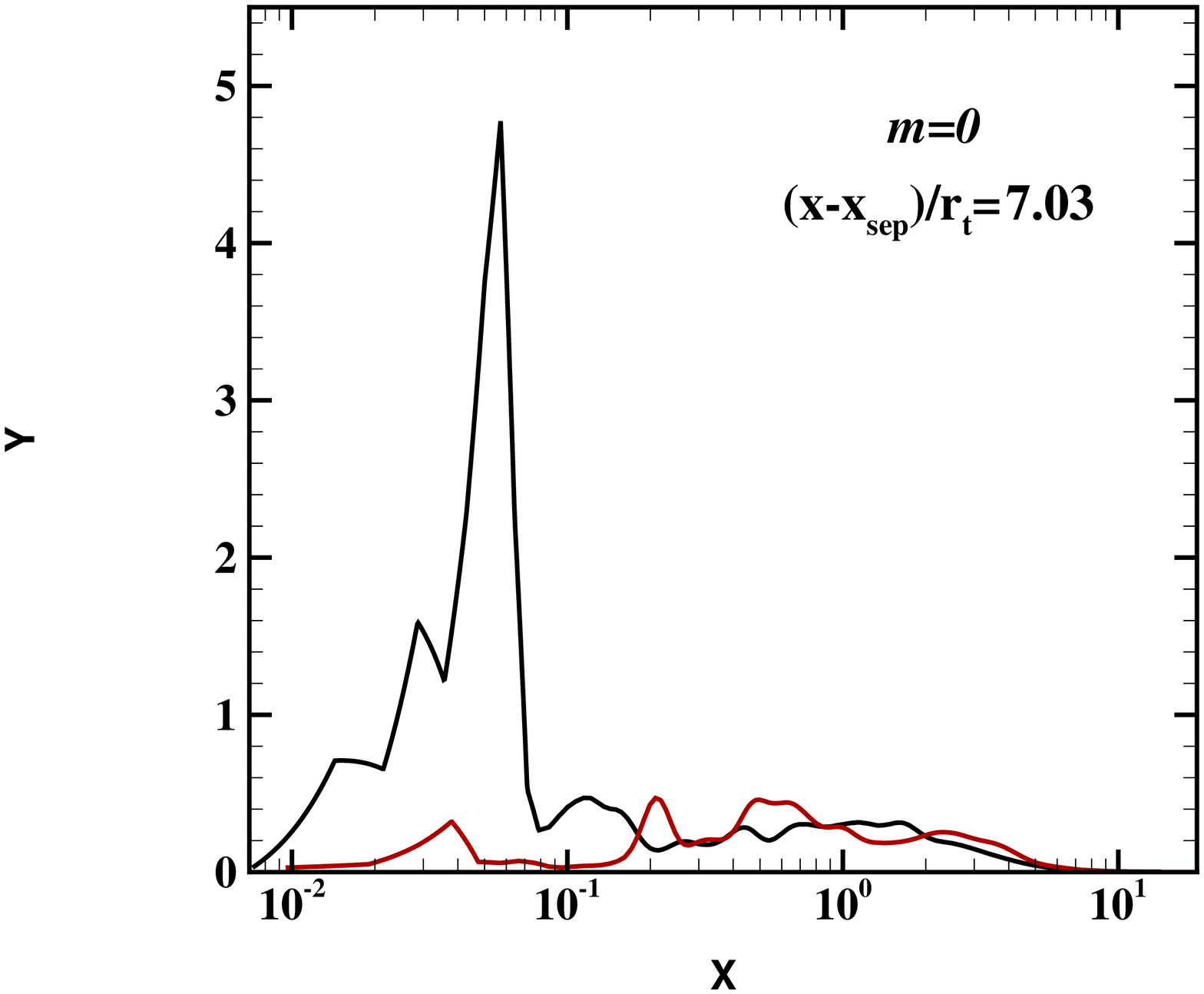} \hskip 1em
(d)\includegraphics[width = 0.3\textwidth,angle=270,clip]{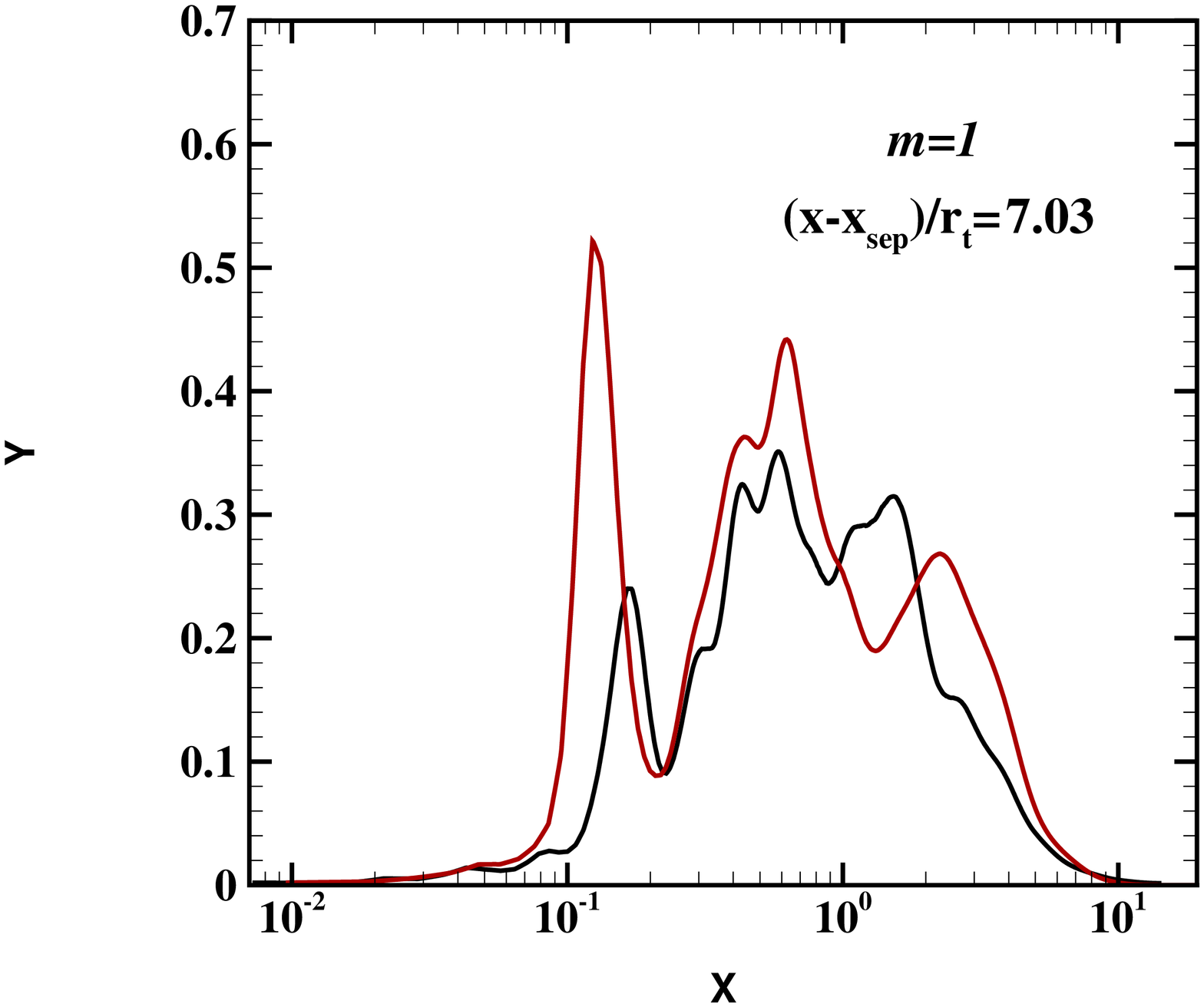} \vskip 1em
        \caption{Premultiplied spectra of the zeroth (left) and first (right) Fourier 
		 azimuthal mode at $x-x_{sep}= 0.16 r_t$ (a,b) and $7.03 r_t$ (c,d). 
		 The black solid line refers to the dual nozzle while the red solid one to the TIC.}
  \label{fig:wavfreq_modes_tic}
\end{figure*}

The proper way to correlate the frequency and energy content of the wall-pressure signature with 
the aerodynamic side loads is to evaluate the Fourier azimuthal wavenumber-frequency spectrum, 
defined as
\begin{equation}
\begin{aligned}
        \phi_{pp} (x,m,f) =
        \int_{-\infty}^{\infty} \int_0^{2 \pi} R_{pp} (x,0,\Delta \theta, \Delta \tau) \, e^{-i (m \Delta \theta + f \Delta \tau)} \mathrm{d}(\Delta \theta) \, \mathrm{d} (\Delta \tau) &,
\end{aligned}
\end{equation}
whit $R_{pp}$ indicating the space-time correlation function of the wall-pressure fluctuations: 
\begin{equation}
        R_{pp}(x,\Delta x,\Delta \theta,\Delta \tau) = \langle{p_w^{'}(x,\theta,t) \, p_w^{'}(x+\Delta x,\theta+\Delta \theta,t+\Delta \tau)}\rangle
\end{equation}
where $\Delta x$ and $\Delta \theta$ are the spatial separations 
in the streamwise and azimuthal directions, $\Delta \tau$ is the time delay, 
and $\langle~\rangle$ denotes averaging with respect to the azimuthal direction (exploiting homogeneity) and time.
The symbol $m$ indicates the mode number in the azimuthal direction, also called Fourier-azimuthal wavenumber and 
it  is noteworthy to remember that the asymmetric mode $m = 1$ is the only one
capable of providing a contribution to the side loads, as also
shown in Fig.\ref{fig:modes_sketch} where the real part ($m>0$) of the first five 
Fourier azimuthal modes is reported.
Following ~\citet{Baars2012}, we introduce  
the total resolved energy $\Lambda(x)$ for each longitudinal station of the nozzle as:
\begin{equation}
\Lambda(x)=\sum_{m}\lambda^{(m)}(x)
\end{equation}
where $\lambda^{(m)}(x)$ is the variance of the $m^{th}$ time-dependent Fourier-azimuthal mode coefficient.
On the base of these definitions it is possible to show in Fig.~\ref{fig:mode_histo} 
the eigenspectra of the Fourier modes as fractions of
the resolved energy per mode $\alpha(x,m)$:
\begin{equation}
\alpha(x,m) = \frac{\lambda^{(m)}(x)}{\Lambda(x)}\, ,
\end{equation}
for two axial stations for both nozzles. 
The axial distances are evaluated from the separation location and 
normalized with throat radius for comparison purposes. 
Starting with the dual-bell nozzle case, Fig.~\ref{fig:mode_histo} (a), 
it is evident that the energy of the zeroth  mode is by far the dominant one along the nozzle. 
The energy of the mode $m=1$ is nearly negligible near the separation location and 
shows an increment at the downstream station, but always remaining a small fraction (1.8\% at most)
of the total resolved energy. The higher modes appear to be non-negligible (but always less than 1\%)
only at the downstream station. The picture highlighted by Fig.~\ref{fig:mode_histo} (b)
for the TIC nozzle is qualitatively similar but with a notable exception. Indeed, it is possible
to appreciate a non-negligible decrease of the energy of the zeroth mode along the nozzle and an
important growth of the first mode, that at $(x-x_{sep})/r_t = 7.03$ provides
a 30\% contribution to the total resolved energy. 
The energy of the higher modes is still negligible near the shock location, giving a minimal contribution only 
at the downstream station. 
To evaluate the frequency content of the time-dependent Fourier azimuthal coefficients, 
their power spectral densities are computed for both nozzles.
Fig.~\ref{fig:wavfreq_modes_tic} shows the spectra for the symmetric ($m=0$)
and asymmetric ($m=1$) modes at the same two axial stations of Fig.~\ref{fig:mode_histo}.
To compare the two different nozzles, a Strouhal ($St$) number is introduced 
as in~\citet{Tam1986}:
\begin{equation}
St = f \frac{D_j}{U_j},
\end{equation}
where $U_j$ is the fully expanded jet velocity
\begin{equation}
U_j = \sqrt{\gamma R T_0}\frac{M_j}{\sqrt{1+\frac{\gamma-1}{2}M_j^2}},
\end{equation}
$T_0$ is the stagnation temperature, $\gamma$ the specific heat ratio, $R$ the
air constant, $M_j$ is the fully adapted Mach number, which is a function of the nozzle
pressure ratio through the isentropic relation.
The length-scale $D_j$ is computed as a function of $M_j$, the design Mach number
$M_d$ and the nozzle exit diameter $D$ through the mass flux conservation
\begin{equation}
\frac{D_j}{D}=\Bigg(\frac{1+\frac{\gamma-1}{2}M_j^2}
{1+\frac{\gamma-1}{2}M_d^2}\Bigg)^{\frac{\gamma+1}{4(\gamma-1)}} \Big(\frac{M_d}{M_j}\Big)^{1/2}.
\end{equation}
Bearing in mind the picture of the wall-pressure power spectral densities (Fig.~\ref{fig:mapspec}),
we see in Fig.~\ref{fig:wavfreq_modes_tic} (a) that the high energy peaks
characterising the shock region, at 833 Hz ($St=$ 0.057) for 
the dual-bell nozzle and at 315 Hz ($St=$ 0.038) for the TIC nozzle, belong to the symmetric mode. 
In the dual-bell case, the zeroth mode maintains its importance along the nozzle, 
while an energy decrease can be observed in the TIC case, Fig.~\ref{fig:wavfreq_modes_tic} (c), 
coherently with the contours in Fig.~\ref{fig:mapspec}. 
As far as the asymmetric mode is concerned, we see in Fig.~\ref{fig:wavfreq_modes_tic} (b) and (d) 
that the dual-bell shows some energy around $\approx$ 2300 Hz ($St=$ 0.16) and that this peak persists downstream.
The same picture appears for the TIC case, but with a much higher energy contribution: near 
the shock it appears a clear tone at approximately 1 kHz ($St=$ 0.12). 
Moving downstream, the peak at $St=$ 0.12 is still the most important but 
now some energy appears, as in the dual-bell case,   
at higher frequencies (around $St \approx$ 0.6), indicating the presence of the asymmetric mode in the 
turbulent shear layer.
As already shown by~\citet{Baars2012},~\citet{Jaunet2017} and \citet{martelli2019flow} it 
seems clear that the flow separation in a TIC nozzle is characterised by a shock 
movement which is mainly symmetric (piston-like oscillation) but with some energy 
present in the helical mode. This last mode, together with the higher ones,
characterises  also the detached shear layer, as shown by its footprint on the wall-pressure signature. 
This scenario is quite different from that observed in the dual-bell case, where
the energy of the helical mode is almost negligible with respect to that of the symmetrical one. 
All the peaks frequencies and $St$ numbers of Fig.~\ref{fig:wavfreq_modes_tic} are
reported in Table~\ref{tab:modespeaks}. It is also adopted a modified Strouhal
number $St^*= f \frac{L}{U_j}$, where L is the distance between the mean separation point
and the nozzle exit plane as defined before.
From the data reported in the table it can be seen that $L$ is, as expected,
a better length scale for the acoustic resonance and in fact, the peak $St^*$'s for
the $m=0$ mode  are very similar for the two nozzles. On the contrary, it seems that
$D_j$ is a better length scale for the helical mode and now the $St$'s are closer. 
\begin{table}
\centering
\caption{Modes peak frequency, St and St$^*$ numbers for Dual-Bell and TIC nozzles.}
\label{tab:modespeaks}
\begin{tabular}{ccccccccccccc}
\hline\noalign{\smallskip}
$(x-x_{sep})/r_t$ & \multicolumn{4}{c}{f (Hz)} & \multicolumn{4}{c}{St} & \multicolumn{4}{c}{St$^*$} \\
\cmidrule(lr){2-5}\cmidrule(lr){6-9}\cmidrule(lr){10-13}
 & \multicolumn{2}{c}{$m=0$} & \multicolumn{2}{c}{$m=1$} & \multicolumn{2}{c}{$m=0$} & \multicolumn{2}{c}{$m=1$} & \multicolumn{2}{c}{$m=0$} & \multicolumn{2}{c}{$m=1$} \\
\cmidrule(lr){2-3}\cmidrule(lr){4-5}\cmidrule(lr){6-7}\cmidrule(lr){8-9}\cmidrule(lr){10-11}\cmidrule(lr){12-13}
 & DB & TIC & DB & TIC & DB & TIC & DB & TIC & DB & TIC & DB & TIC\\
\noalign{\smallskip}\hline\noalign{\smallskip}
0.16 & 830 & 315 & 2330 & 996 & 0.057 & 0.038 & 0.16 & 0.12 & 0.10 & 0.13 & 0.29 & 0.40\\
7.03 & 830 & 315 & 2330 & 996 & 0.057 & 0.038 & 0.16 & 0.12 & 0.10 & 0.13 & 0.29 & 0.40\\
\noalign{\smallskip}\hline
\end{tabular}
\end{table}

The very low energy content of the first azimuthal mode in the dual-bell 
nozzle, with respect to the TIC case, is an indication that 
the dual bell working in sea-level 
mode at the present NPR should not develop 
a significant level of side loads.
\citet{Jaunet2017} and \citet{martelli2019flow} speculated the existence of
a screech-like mechanism inside conventional nozzles, 
with free shock separation, associated to this helical mode. 
\citet{martelli2019flow} proposed a path for the feedback loop, involving 
the downstream propagation of hydrodynamic instabilities along the shear layers 
and in the nozzle core downstream the Mach disk and the upstream travelling of 
the acoustic waves, generated by the interaction 
of the vortices with the shock cells, in the subsonic turbulent recirculating region 
inside the nozzle.
Indeed, according to~\citet{powell1953mechanism,powell1953noise}, 
every aeroacoustic resonance in high-speed jets can be decomposed into four 
main processes: i) the downstream propagation of 
energy through hydrodynamic instabilities, ii) an acoustic generation process in which the 
downstream perturbations are converted into upstream perturbations, iii) the upstream 
propagation of these disturbances, iv) the generation of new hydrodynamic instabilities through the
forcing of a sensitive point excited by the upstream acoustic waves (the nozzle
lip in external screech). The last process is generally called receptivity process~\citep{edgington2019}.
One of the criterion that Powell established for the sustainability of a feedback loop
of this type is the following:
\begin{equation}
\centering
	q_d \eta_g \eta_u \eta_r \ge 1.
\label{eq:powell}
\end{equation}
where $q_d$ is the gain associated with the downstream process, $\eta_g$ is the conversion
efficiency of the hydrodynamic disturbances into acoustic waves, $\eta_u$ is the efficiency
of the upstream disturbances transmission and $\eta_r$ is the efficiency of the
receptivity process.
In the TIC nozzle, this process involves the forcing of the separation line and 
of the separation shock, which induces the generation of a new hydrodynamic instability. 
In the dual-bell nozzle the helical mode is only slightly excited
due to the presence of the inflection point that anchors the separation shock 
and constrains its movement. In particular, 
the inflection point induces the shock to move symmetrically, thus
modifying the efficiency of the receptivity process ($\eta_r$) of the feedback loop,
which hampers the generation of the first azimuthal mode.
A quantitative analysis and comparison of the side-loads generation in the dual-bell and TIC nozzles
is reported in the following section.

\subsection{Aerodynamic loads distribution}
\label{sec:sl}

%
\begin{figure*}
  \centering
         \psfrag{X}[c][c][1.2]{$F_z/P_c A_t$}
         \psfrag{Y}[c][c][1.2]{$F_y/P_c A_t$}
(a)\includegraphics[width = 0.3\textwidth,angle=270,clip]{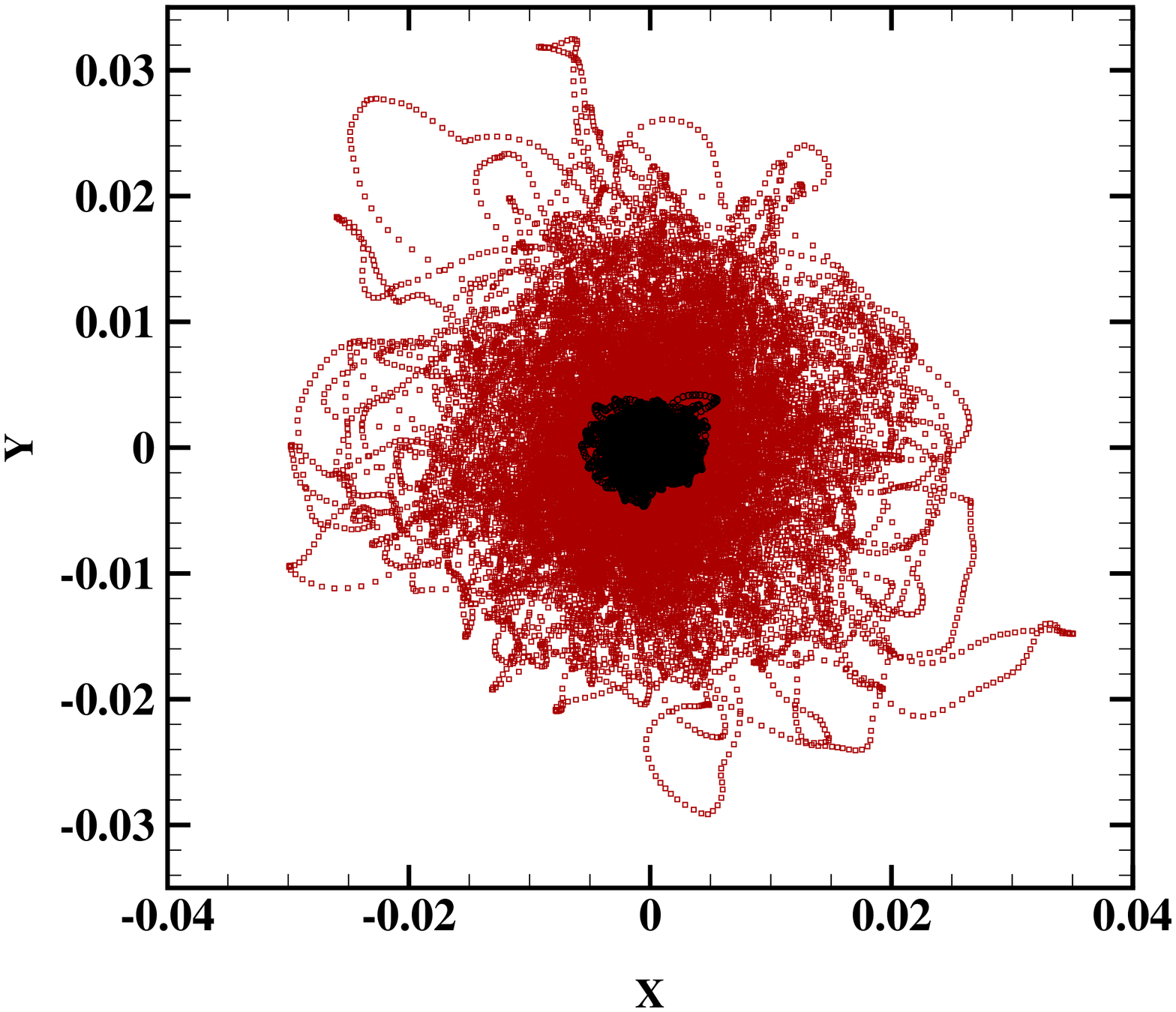} \hskip 1em
        \psfrag{X}[c][c][1.2]{$t\,(s)$}
         \psfrag{Y}[b][ ][1.2]{$|F|/P_c A_t$}
(b)\includegraphics[width = 0.3\textwidth,angle=270,clip]{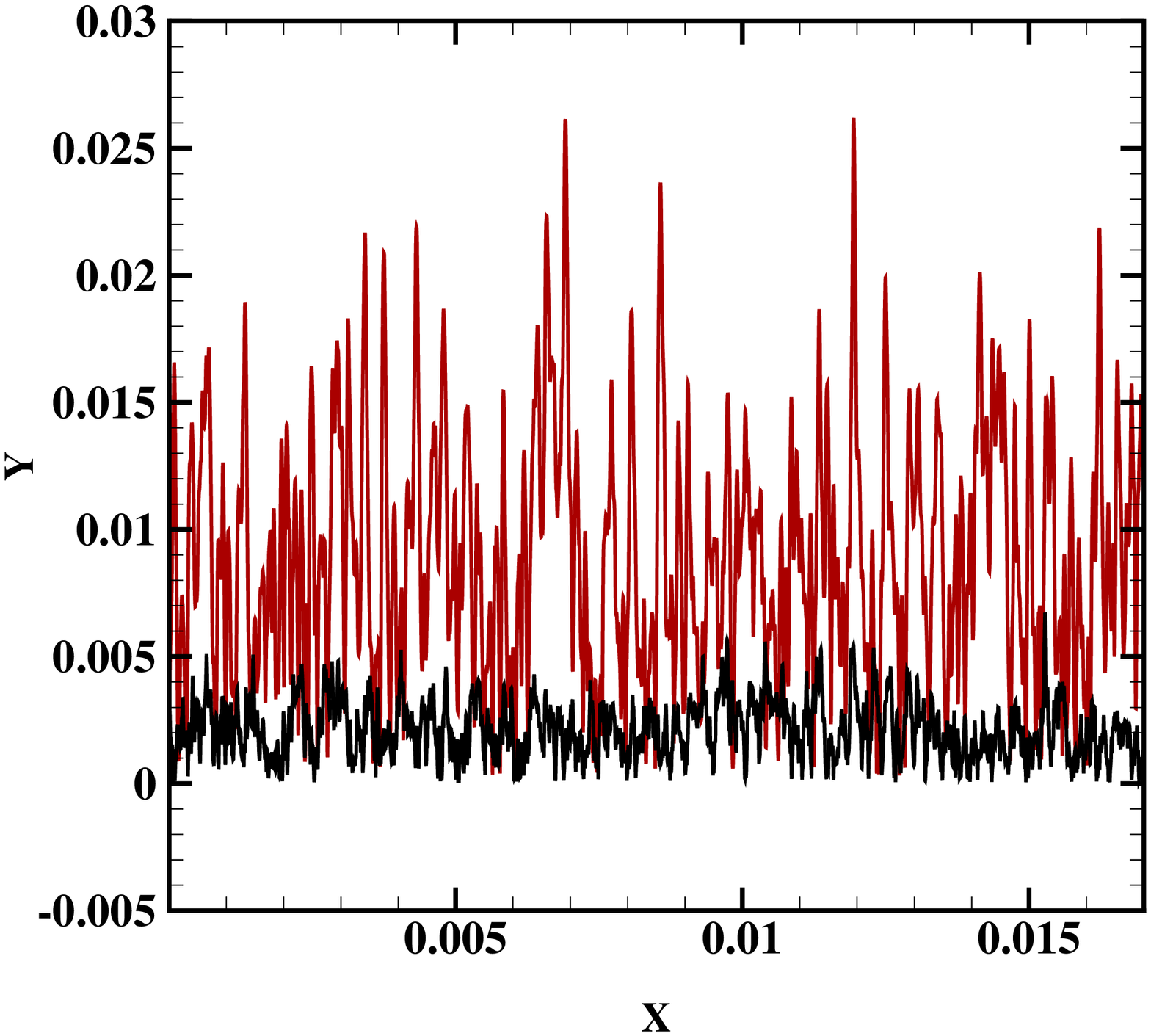} \hskip 1em
        \caption{Comparison between dual-bell and TIC nozzles of the time distribution of the 
	         side-loads components $F_y$ and $F_z$ (a) and magnitude $|F|$ (b). 
	         The black symbols and solid line refer to the dual-bell nozzle while the red 
		 symbols and solid line to the TIC.}
  \label{fig:sl_tic_db}
\end{figure*}

\begin{figure}
  \centering
(a)\includegraphics[width = 0.28\textwidth,angle=270,clip]{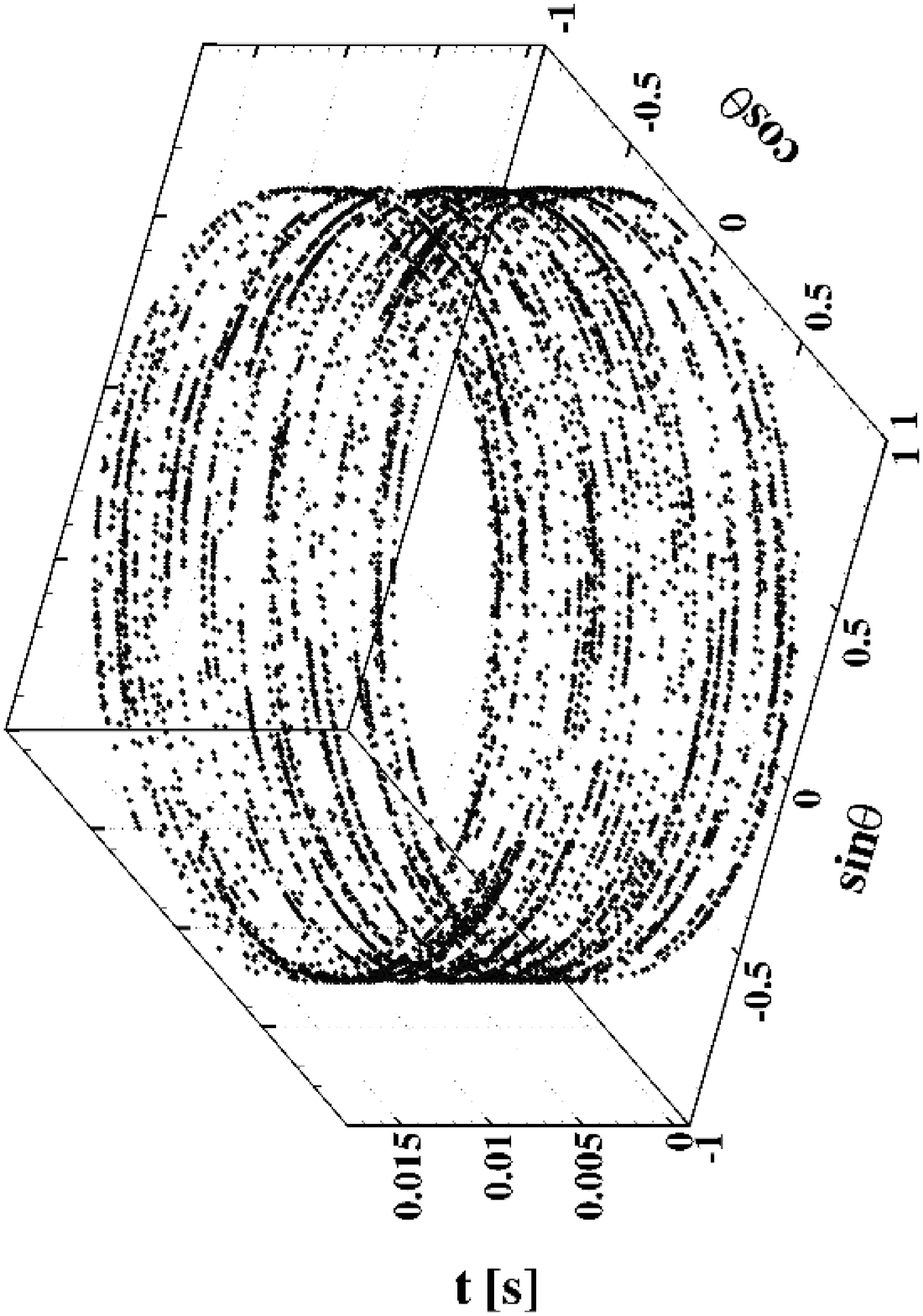} \hskip 1em
(b)\includegraphics[width = 0.28\textwidth,angle=270,clip]{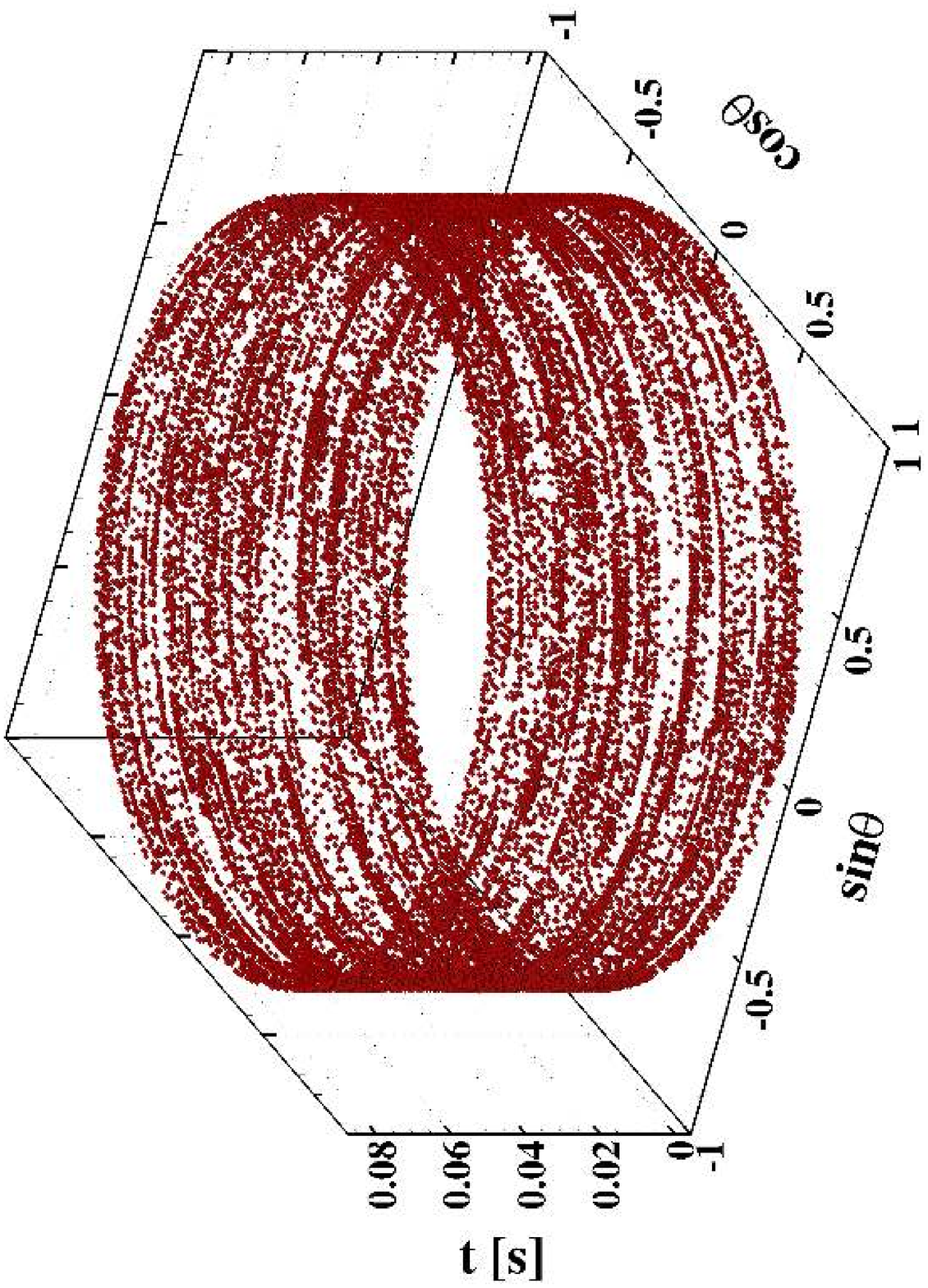} \hskip 1em
  \caption{Time history of side-loads direction: (a) Dual-Bell and (b) TIC nozzle.}
  \label{fig:sl_dir}
\end{figure}

\begin{figure*}
  \centering
	 \psfrag{X}[t][ ][1.2]{$F_y/\sigma_{F_y}$}
         \psfrag{Y}[b][ ][1.2]{$pdf$}
(a)\includegraphics[width = 0.21\textwidth,angle=270,clip]{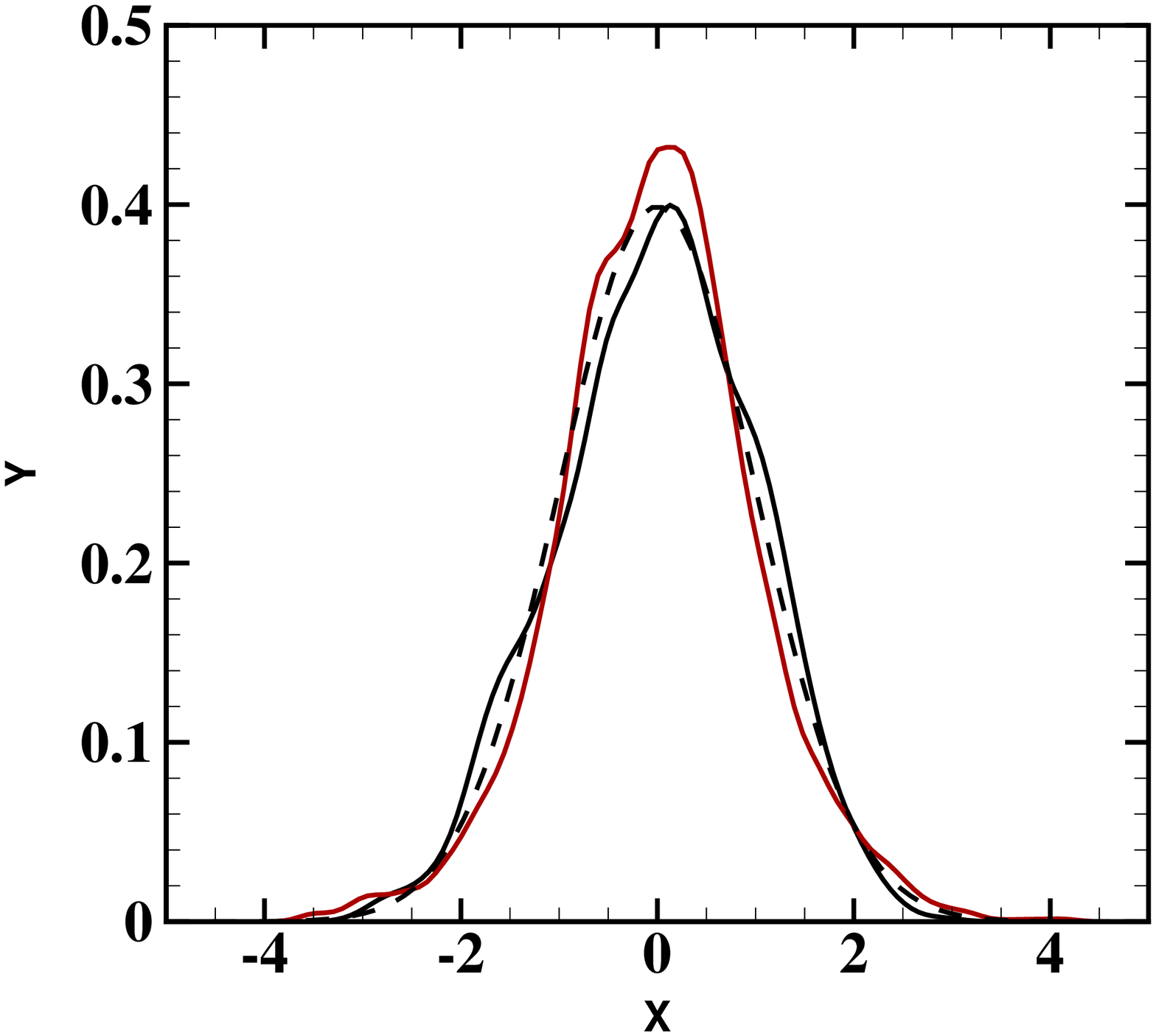} 
	 \psfrag{X}[t][ ][1.2]{$F_z/\sigma_{F_z}$}
         \psfrag{Y}[b][ ][1.2]{$pdf$}
(b)\includegraphics[width = 0.21\textwidth,angle=270,clip]{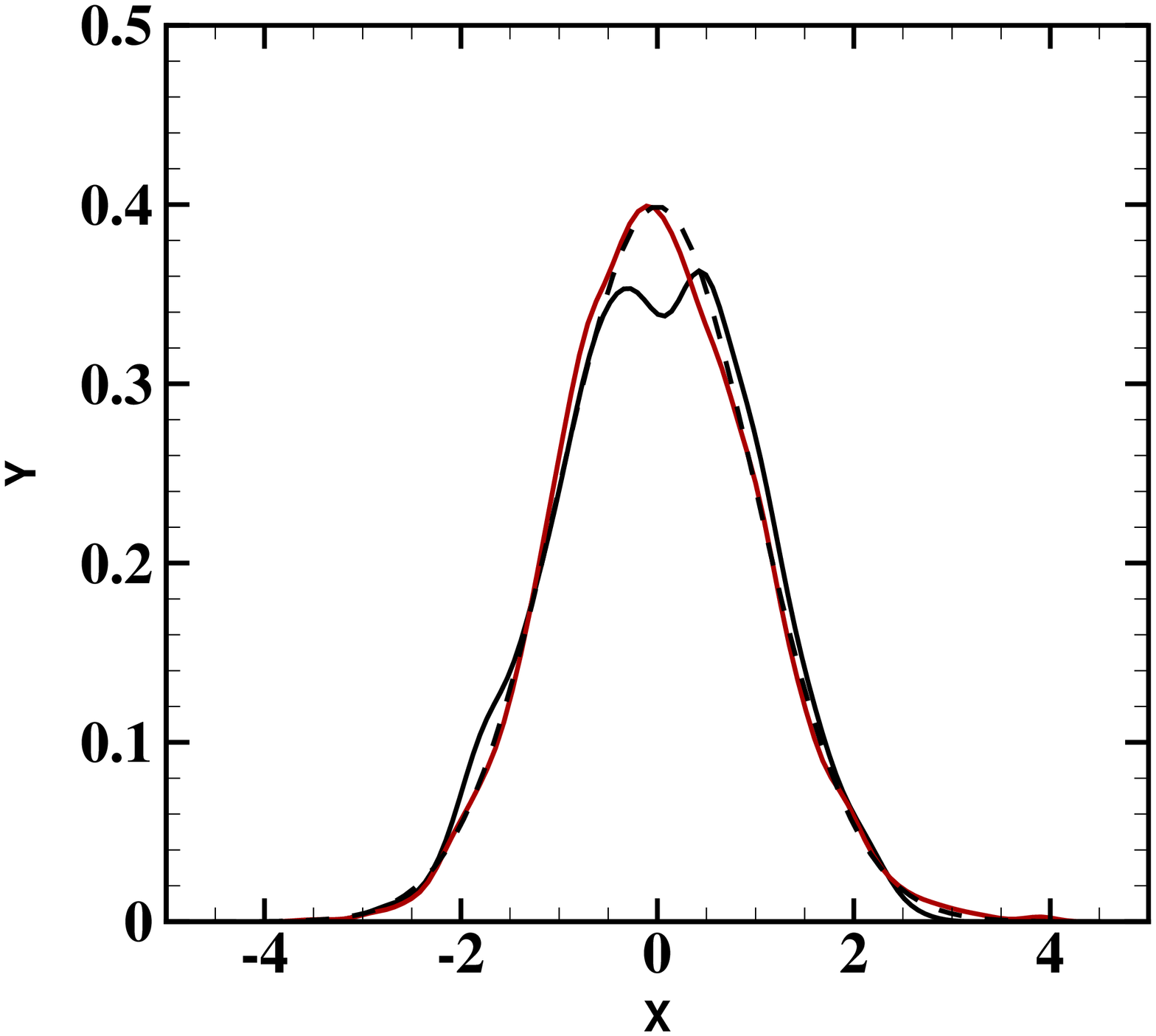} 
	 \psfrag{X}[t][ ][1.2]{$|F|/\sigma_{|F|}$}
         \psfrag{Y}[b][ ][1.2]{$pdf$}
(c)\includegraphics[width = 0.21\textwidth,angle=270,clip]{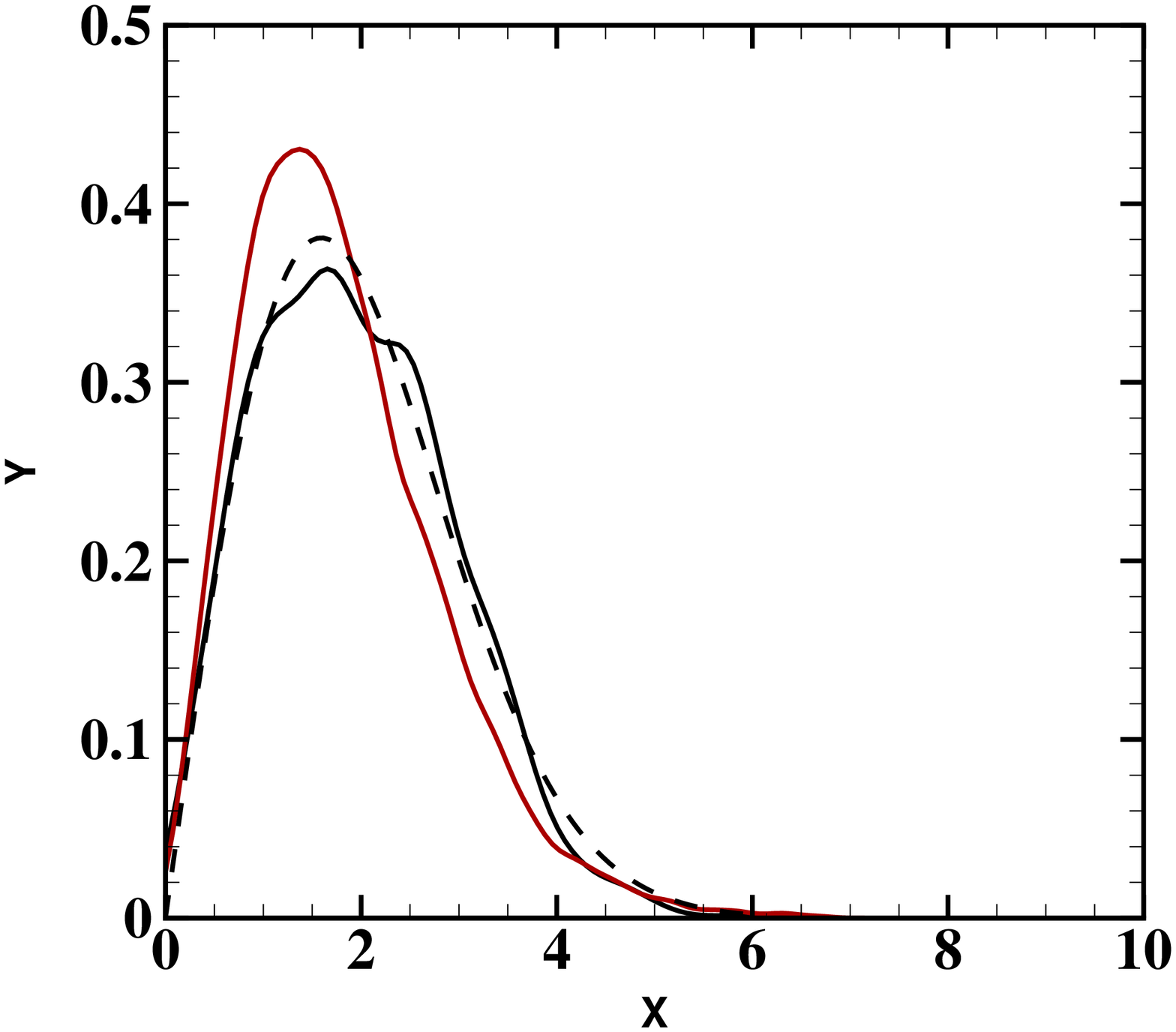} \vskip 1em
        \caption{Comparison between the dual-bell and TIC nozzles of the probability density function of the 
	         side-loads components (a,b) and magnitude (c).
	         Black solid lines represent the dual-bell nozzle, the red solid lines represent the TIC nozzle while the
		 black dashed line is the theoretical distribution.}
  \label{fig:sl_pdf}
\end{figure*}

The wall-pressure distribution is integrated along the nozzle wall (longitudinal and azimuthal direction)
to obtain the time distribution of the side-loads components ($F_y$ and $F_z$) and magnitude 
($|F|=\sqrt{F_y^2+F_z^2}$), which are reported in Fig.~\ref{fig:sl_tic_db}.
The thrust and side-loads components are calculated by integration of the wall-pressure field as
\begin{equation}
F_x(t) = \int_{S_{nozzle}} [p_a-p_{w}(t,x,\theta)]\cdot n_x \,dS,
\end{equation}
\begin{equation}
F_y(t) = \int_{S_{nozzle}} [p_a-p_{w}(t,x,\theta)]\cdot n_y \,dS,
\end{equation}
\begin{equation}
F_z(t) = \int_{S_{nozzle}} [p_a-p_{w}(t,x,\theta)]\cdot n_z \,dS.
\end{equation}
The figures highlight that the side-loads components oscillate around the zero mean value
and have a randomic distribution in time. Consistently with the previous discussion
on the energy of the asymmetric $m=1$ mode, the magnitude of the aerodynamic side loads in the TIC nozzle
is found to be remarkably higher than that observed in the dual-bell nozzle.
The experimental side-loads time history is not available from~\citet{verma2015},
therefore, the numerical results are compared with the
experimental data of \citet{geninstark2011}, where a test campaign
on a similar subscale dual-bell nozzle (named DB2) was performed. This nozzle has a slightly
different area ratio of the second bell ($\epsilon_e = 24$) and a different inflection angle
($\alpha_i = 5^{\circ}$). The simulation results show an averaged value of the side loads of
$3.26$ N,  which well agrees with
the value of $3.42$ N obtained by the experiments.
The time evolution of the side-loads vector direction ($cos(\theta)$, $sin(\theta)$) 
with respect to the z axis, is 
reported in Fig.~\ref{fig:sl_dir} as a function of time.
This figure shows that the side loads have not a preferential direction in the space, similarly 
to the results obtained by~\citet{Deck2002}, who performed a numerical campaign to predict 
side loads in a sub-scale TIC nozzle. 
The probability density functions of $F_y(t)$, $F_z(t)$ and $|F|(t)$ 
are shown in Fig.~\ref{fig:sl_pdf} (a), (b) and (c) respectively for both nozzles.
It clearly appears that the side-loads components
are two independent normal random variables with zero mean value and the same standard deviation,
thus indicating that their distribution is Gaussian. Indeed, the numerical distribution well fit
the theoretical Gaussian one. As a consequence, the probability distribution of $|F|$, 
 Fig.~\ref{fig:sl_pdf}(c), is a Rayleigh distribution, as shown by \citet{Dumnov1996}, 
who studied the experimental statistical distribution of side loads in a sub-scale nozzle.
\begin{figure}
  \centering
  \psfrag{X}[c][c][1.2]{$x/r_t$}
  \psfrag{Y}[b][ ][1.2]{$|F|/P_c A_t$}
(a)\includegraphics[width = 0.3\textwidth,angle=270,clip]{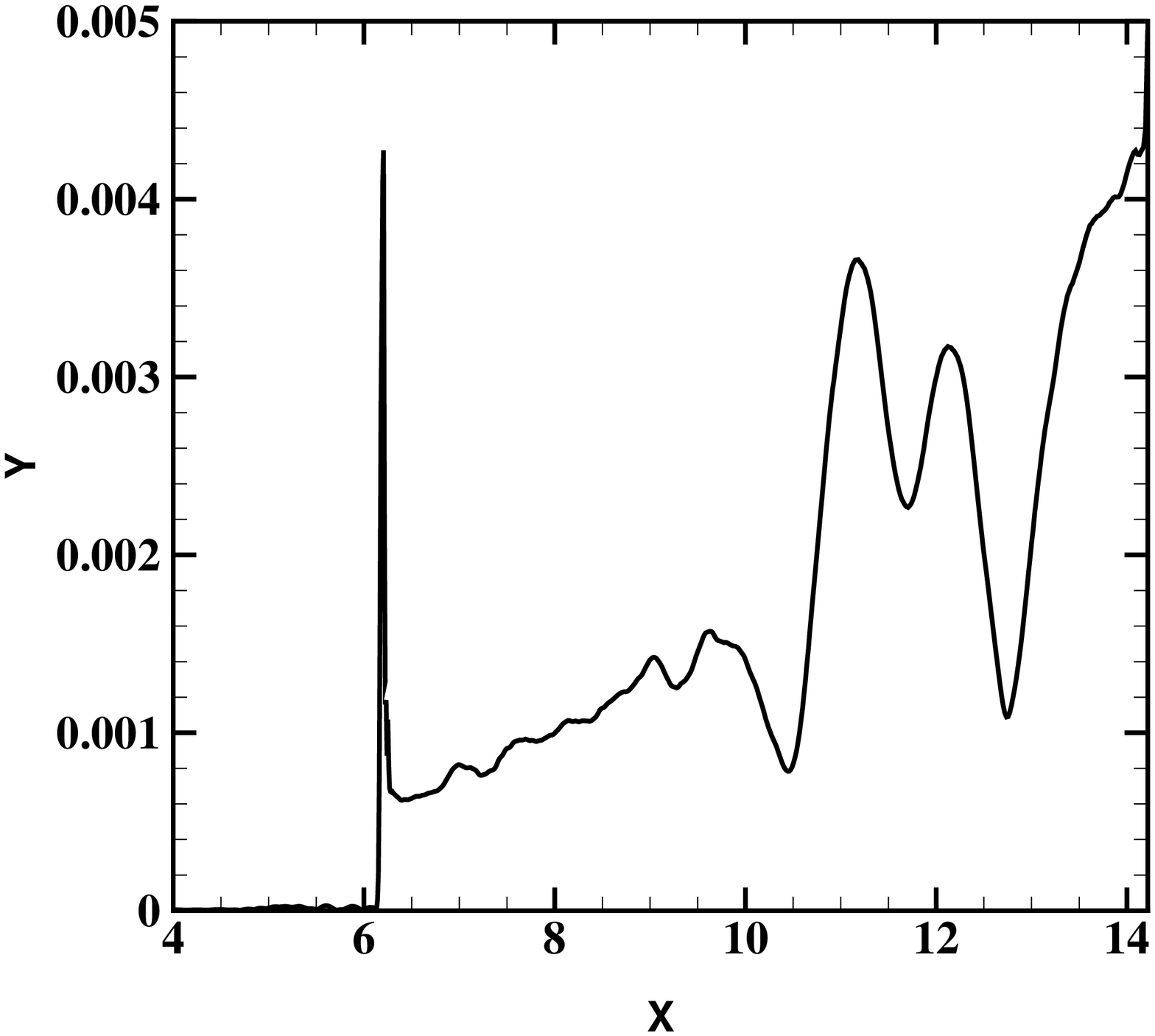} \hskip 1em
  \psfrag{X}[c][c][1.2]{$x/r_t$}
  \psfrag{Y}[c][c][1.2]{$|F|/P_c A_t$}
(b)\includegraphics[width = 0.3\textwidth,angle=270,clip]{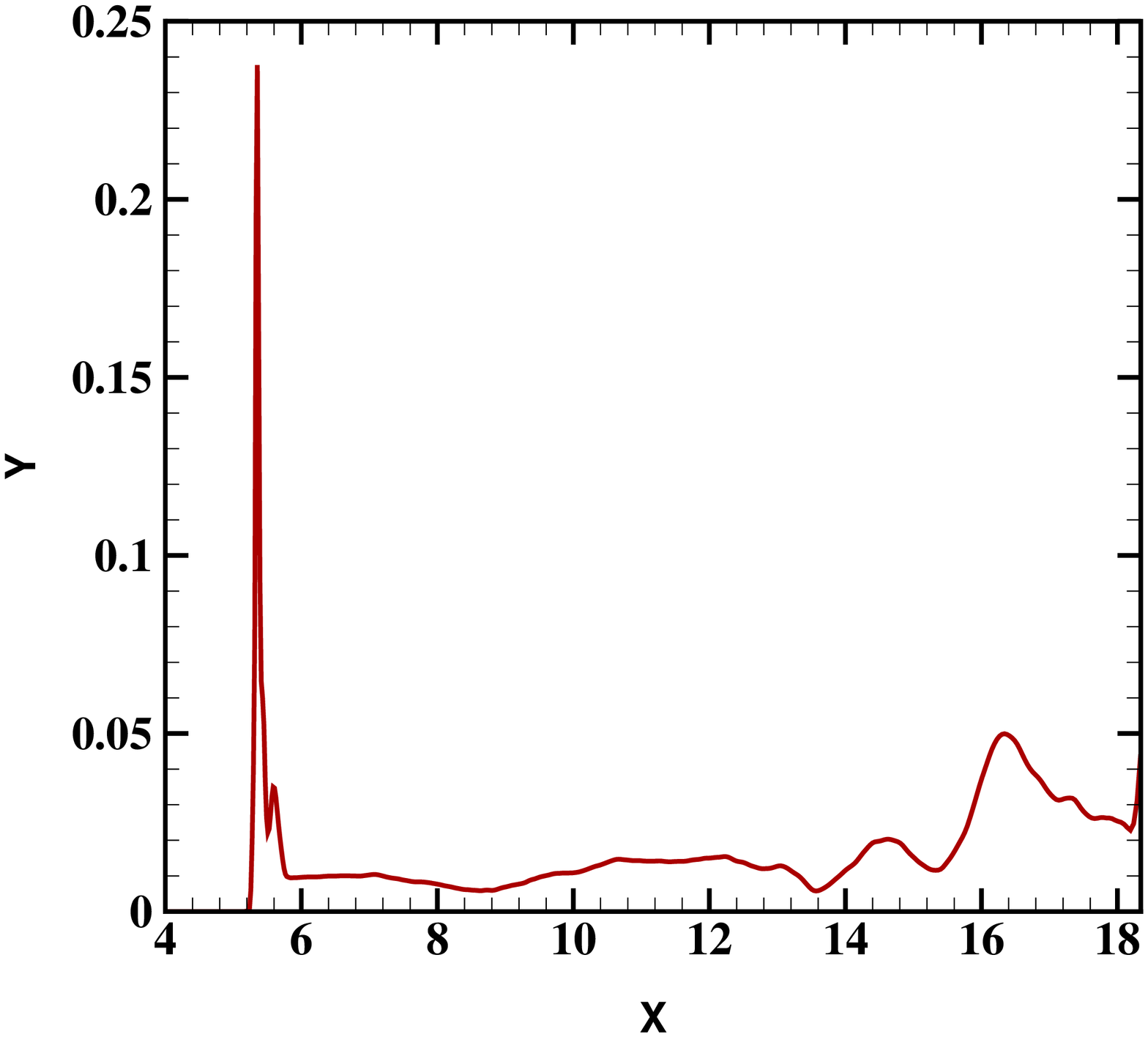} \hskip 1em
  \caption{Evolution along the x-axis of the side-loads magnitude averaged in time: (a) Dual-Bell 
           nozzle and (b) TIC nozzle.}
  \label{fig:sl_axis}
\end{figure}

\begin{figure}
  \centering
  \psfrag{X}[c][c][1.2]{$t\,(s)$}
  \psfrag{Y}[c][c][1.2]{$F_x/P_0 A_t$}
(a)  \includegraphics[width = 0.3\textwidth,angle=270,clip]{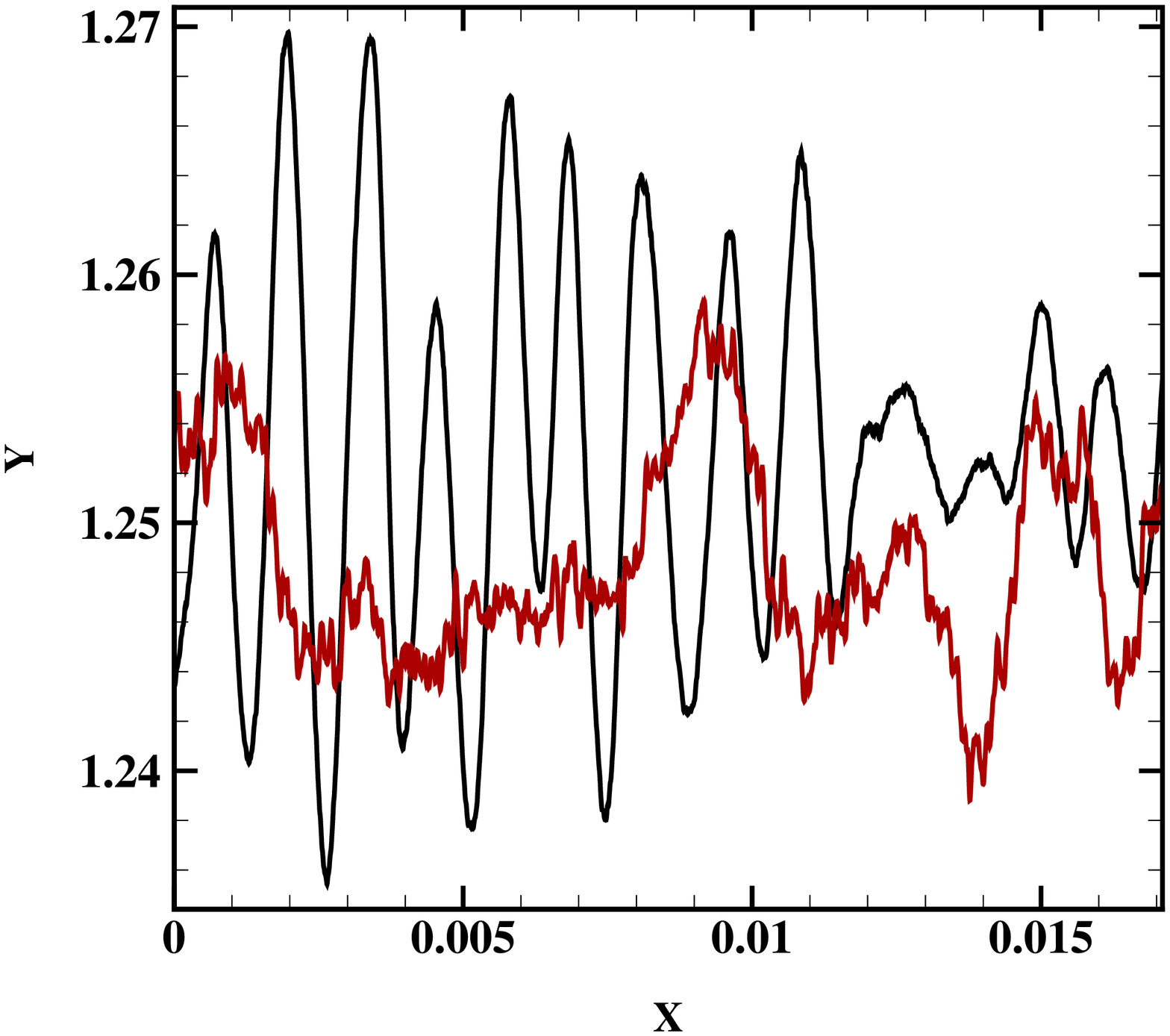} \hskip 1em
  \psfrag{X}[c][c][1.2]{$St^*$}
  \psfrag{Y}[c][c][1.2]{$G(f)\,f/\sigma^2$}
(b)   \includegraphics[width = 0.3\textwidth,angle=270,clip]{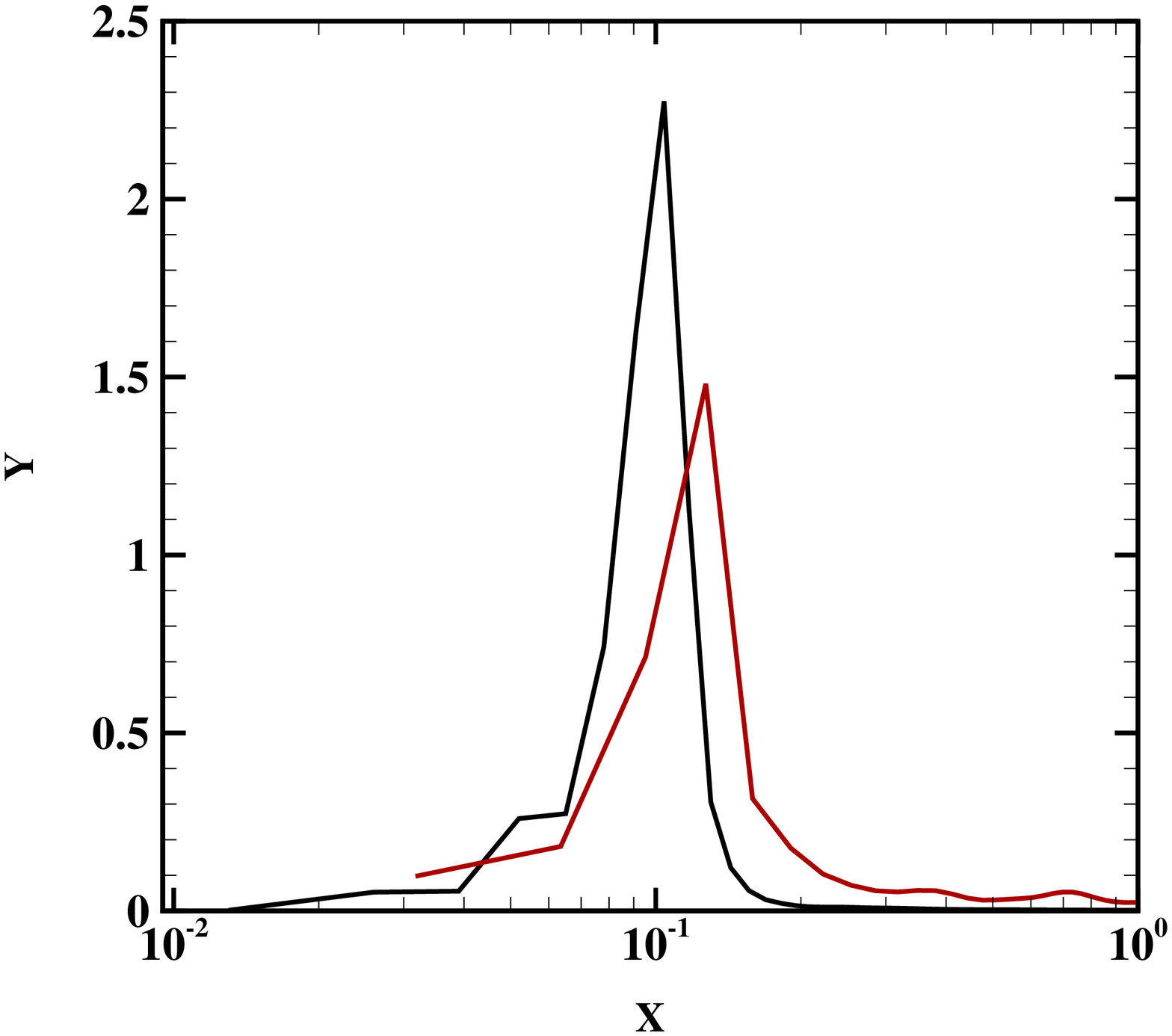} \vskip 1em
        \caption{(a) Time history of the thrust coefficient $F_x/(p_0A_t)$ for the dual-bell and TIC nozzles and
                 (b) corresponding premultiplied power spectral densities. 
                The black solid line denotes the dual-bell nozzle, the red solid line refers to the TIC nozzle.}
  \label{fig:thrust_tic}
\end{figure}

To better characterize the origin of the aerodynamic side loads, and identify the nozzle regions
that most contribute to their generation, the axial distribution of the side-loads magnitude averaged in time 
is shown in Fig.~\ref{fig:sl_axis} for the dual-bell and the TIC geometry.
These profiles have been obtained by integrating, at each streamwise location,
the wall-pressure distribution in the azimuthal direction and averaging in time.
As far as the dual bell is concerned, Fig.~\ref{fig:sl_axis}(a) shows  
a peak in correspondence of the separation-shock location,
followed by a gradual increase until the end of the second bell,
associated with the development of the turbulent shear layer.
This behaviour is in close agreement with the observations made
on the spatial distribution of the first Fourier azimuthal mode. 
The axial distribution of the side-loads in the TIC nozzle, 
Fig.~\ref{fig:sl_axis} (b), further confirms the larger intensity of the
lateral force for this type of nozzle.
In this case, a strong peak is found in the proximity of the shock location,
whereas a flat contribution is observed in the divergent section of the nozzle.
It is worth to notice that these trends are qualitatively similar to that obtained by~\citet{Deck2002}. 
\begin{figure*}
  \centering
         \psfrag{X}[c][c][1.2]{$St$}
         \psfrag{Y}[c][c][1.2]{$G(f)\,f/\sigma^2$}
(a)\includegraphics[width = 0.3\textwidth,angle=270,clip]{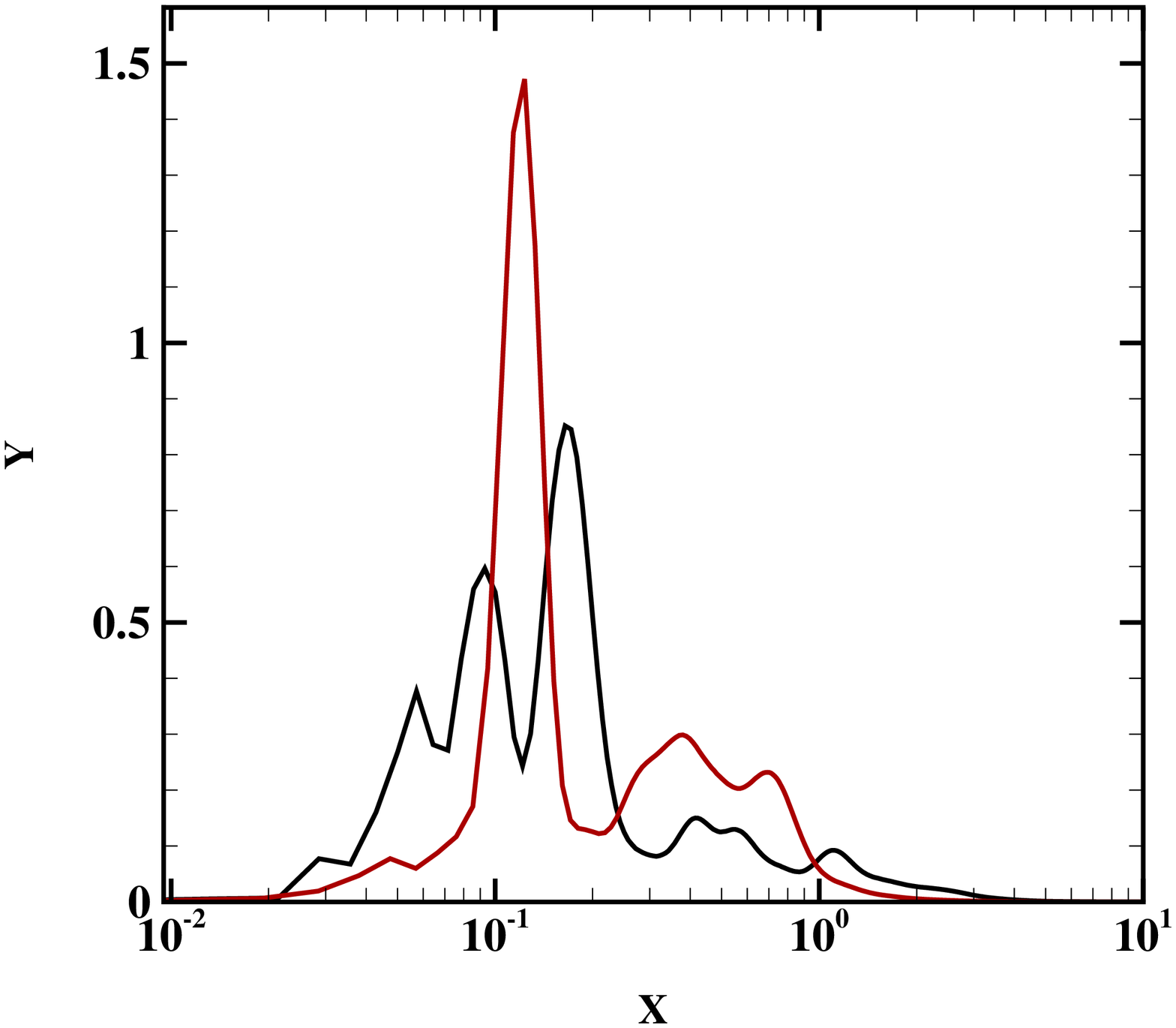} \hskip 1em
(b)\includegraphics[width = 0.3\textwidth,angle=270,clip]{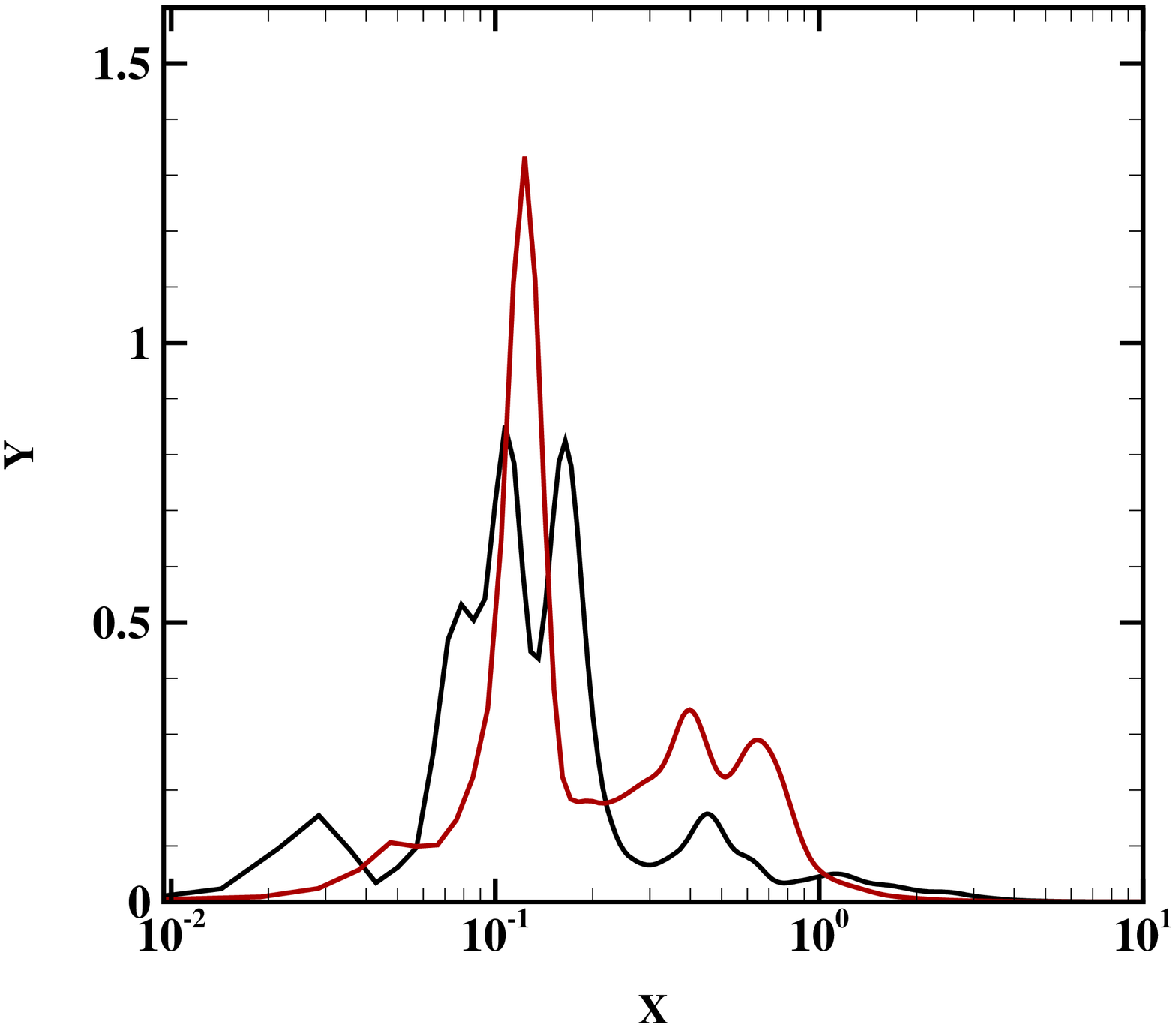} \hskip 1em
        \caption{Normalized premultiplied power spectral densities of the side-loads
                 components $F_y$ (a) and $F_z$ (b). The black solid line refers to the
                 DB nozzle while the red solid one to the TIC.}
  \label{fig:slspec}
\end{figure*}
The Fourier spectra of the lateral forces $F_y$ and $F_z$
for both nozzles are reported in Fig.~\ref{fig:slspec}.
It is clearly visible that the spectra are characterised by two peaks. 
The  Strouhal number corresponding to the higher peak 
is equal to 0.12 for the TIC nozzle and it corresponds, as 
expected, to the peak frequency of the first Fourier azimuthal mode 
(see Table~\ref{tab:modespeaks}). 
The second bump  at higher frequencies corresponds 
to the contribution coming from the turbulent structures 
of the shear layer. The same behaviour characterises the spectra 
of the dual-bell loads components: a major contribution 
coming from the helical mode (peak at $St$=0.16) and 
a second minor contribution from the separated region.
Finally, the dimensionless longitudinal component of the force $F_x/(p_c A_t)$, corresponding
to the thrust coefficient,  
is reported in Fig.~\ref{fig:thrust_tic}(a) for both geometries. 
The two nozzles are characterised by similar levels of thrust oscillation and 
the spectra reported in Fig.~\ref{fig:thrust_tic} (b) show  
that all the energy is concentrated 
at the frequency corresponding to the acoustic resonance given the 
by the piston-like shock movement, $St^*$=0.10 and 0.13 for the dual bell 
and TIC nozzle, respectively.  
This implies that, in the design phase, the level of vibrations
at lift-off and during the first part of the trajectory 
(first operating mode), that are detrimental for the engine structure and the payload, must be carefully addressed.



\section{Conclusions}
\label{sec:conclusions}

A delayed detached eddy simulation (DDES) of a dual-bell nozzle
working in an over-expanded condition with flow separation
anchored at the wall-inflection point has been carried out.
The main goal of the investigation was to figure out the effectiveness
of the wall discontinuity as a flow-separation control device. The study
has been conducted by analyzing the spectral content of the wall-pressure signature,
with the purpose of evaluating the aerodynamic loads and comparing them with those
experienced by a truncated ideal contour (TIC) nozzle~\citep{martelli2019flow}.
The latter is a well-known conventional nozzle, that suffers the onset of
side loads when operates with an internal flow separation.
The dual-bell nozzle geometry has been specified following
the experimental work of~\citet{verma2015}, whose data have been used
to assess the accuracy of the simulation.
The analysis of the unsteady wall-pressure signals showed a good agreement of the
mean wall pressure and the standard deviation of the
pressure fluctuations along the nozzle with the experimental data,
although the standard deviation peak at the shock location seemed to be over-predicted.
The spectral analysis highlighted the presence of an acoustic tone at $\approx$0.8 kHz,
associated with the shock motion. This tone appeared to persist all along the
nozzle wall, as also confirmed by the experimental trend, and it was found to be
an acoustic resonance, its frequency being associated with a one-quarter standing
wave that develops between the mean separation shock location and the nozzle lip.
Also the TIC nozzle was characterised by a strong tone in the low frequency range, even
if its energy decreases along the nozzle wall. Both nozzles also showed the presence of
oscillation energy in the high frequency range (between 10 kHz and 100 kHz), originated
by the turbulent detached shear layer.

The analysis carried out in the wavenumber-frequency space revealed that the
low frequency peaks of both geometries are associated to the zeroth azimuthal Fourier mode,
that is symmetric, and its energy represents the dominant contribution to the total
fluctuation energy. The second important contribution came from the first (helical) mode, which
is the only one to be non-symmetric and capable to trigger lateral forces. In particular,
the TIC nozzle was characterised by a tone in the intermediate frequency range,
($\approx$ 1 kHz, $St$=0.12),
as already shown in literature~\citep{Baars2012,Jaunet2017,martelli2019flow}.
The first Fourier azimuthal mode is instead only slightly excited in the dual-bell nozzle
with a peak at approximately 2300 Hz ($St$=0.16).
Its energy is almost negligible with respect to the zeroth mode and much lower with respect
to the TIC case. Therefore the triggering of the helical mode seems
to be suppressed by the presence of the inflection point, which
mainly forces a symmetric shock movement, altering the receptivity process of the
separation line to the upstream travelling acoustic disturbances.
The evaluation of the aerodynamic lateral forces revealed a qualitatively similar
distribution for both geometries, in particular they follow the Rayleigh distribution
as found by~\citet{Dumnov1996}. The mean magnitude of the lateral force in the dual bell well
agrees with the experimental value of~\citet{geninstark2011} and it is
an order of magnitude lower than the side loads found in the TIC nozzle.
On the other hand, the fluctuation of the thrust coefficient is similar in both
cases and of the order of 1\%. This aspect should be considered in the design phase
and carefully evaluated to avoid annoying vibrations on the payload and engine structure.

This study was funded by Ministero Istruzione Università e Ricerca (grant number RBSI14TKWU, SIR programme 2014). 
The authors declare that they have no conflict of interest.

\end{document}